\documentclass[12pt,sort&compress]{article}
\usepackage{ae,aecompl}
\usepackage[T1]{fontenc}
\usepackage[latin9]{inputenc}
\usepackage{amsmath}
\usepackage{amssymb}
\usepackage{graphicx}
\usepackage{wasysym}
\usepackage[numbers]{natbib}

\makeatletter

\providecommand{\tabularnewline}{\\}

\newcommand{\lyxaddress}[1]{
	\par {\raggedright #1
	\vspace{1.4em}
	\noindent\par}
}

\@ifundefined{date}{}{\date{}}

\usepackage{indentfirst}
\usepackage{amsfonts}
\usepackage{aecompl}\usepackage{sidecap}
\usepackage[section]{placeins}
\usepackage{epsf}

\usepackage{fancyhdr}

\makeatother

\begin{document}
\title{{\Large\textbf{On a relationship between grain boundary free energy,
grain boundary segregation, and grain boundary diffusion}}}
\author{Yuri Mishin\thanks{Corresponding author. Email: ymishin@gmu.edu (Y. Mishin)}}
\maketitle

\lyxaddress{Department of Physics and Astronomy, George Mason University, Fairfax,
VA, USA}
\begin{abstract}
We present a detailed analysis of the universal relationship between
grain boundary (GB) free energy and GB self-diffusion coefficient
derived by Borisov et al.~(1964). This relationship was expressed
by a simple equation that was used in many publications to predict
GB energies on the basis of experimental diffusion data. Meanwhile,
the physical assumptions and approximations underlying the Borisov
model are poorly understood. As a result, the Borisov equation was
often used outside its intended limits. Here, we re-derive the Borisov
equation from ground up, identifying its underlying assumptions, correcting
some errors and inconsistencies, and extending the original model
to the case of impurity diffusion and diffusion by the interstitialcy
and interstitial-dumbbell mechanisms. The meaning of the key assumption
of the Borisov model, related to the free energy of the activated
complex, is discussed, and ways to test this assumption are proposed. 
\end{abstract}

\section{Introduction}

In 1964, Borisov et al.~\citep{Borisov} published a theoretical
model predicting a functional relationship between grain boundary
(GB) energy and the ratio of self-diffusion coefficients in the GB
and in the surrounding lattice. This relationship was expressed by
a simple equation that was subsequently used by many authors to calculate
GB energies based on experimental or simulated diffusion data.

Some applications of the Borisov model\footnote{For brevity, we occasionally refer to the model proposed by Borisov
et al.~\citep{Borisov} as the ``Borisov model''. This term does
not imply any attribution of credit among the authors. Most likely,
the list of the authors is simply alphabetical, as was common in the
1960s.} produced meaningful and fairly consistent results. In some cases,
they yielded significant insights into GB properties, such as the
temperature dependence of the GB free energy and its decrease due
to impurity segregation. In other cases, the model was used beyond
its intended limits because the authors did not fully understand its
underlying assumptions. Indeed, the original publication \citep{Borisov}
offered only a brief derivation of the main equation without justifying
or even clearly stating the simplifying assumptions and approximations.
Many subsequent publications uncritically repeated this equation,
sometimes rewriting it in different mathematical forms, without re-deriving
it or analyzing its foundations.

Our goal is to critically reexamine the Borisov model to better understand
its assumptions and limitations. We present a careful step-by-step
derivation of the main equation of the model and generalize it to
impurity diffusion and to diffusion mechanisms not considered in the
original publication \citep{Borisov}. We review recent applications
of the model and suggest experimental and computational approaches
that could leverage the capabilities of the model and test some of
its most fundamental assumptions.

We start by providing a theoretical background for the subsequent
discussion by reviewing the statistical mechanics of point defects
and diffusion in solids (section \ref{sec:Background}). This material
is well-known, but scattered over many original and review articles
and textbooks. We find it useful to briefly summarize this material
in one place and introduce the relevant terminology and notation that
will be used throughout the rest of this article. We also discuss
the meaning of the activated complex and the definition of its free
energy (section \ref{subsec:activated_complex}), which are the key
concepts in the Borisov model. In section \ref{sec:Borisov}, we formulate
the main assumptions that underpin the Borisov model and derive the
relationships between the GB free energy and the coefficients of self-diffusion
and impurity diffusion by the vacancy and interstitial (direct and
indirect) mechanisms. The derivations reveal another important parameter
of the model that was not mentioned by Borisov et al.~\citep{Borisov},
which is the number of atoms comprising the activated complex. In
section \ref{sec:Discussion}, we discuss implications of the derived
equations and review the recent literature in which the Borisov model
was used to predict GB energies. The discussion also includes the
possible future tests of the activated complex assumption by atomistic
computer simulations. Section \ref{sec:Summary} presents a brief
summary of this work.

In deriving the equations, we avoid using the notion of the free energy
of an atom. The latter is not well-defined, except when using interatomic
potentials as a model of atomic interactions or within the CALPHAD
framework. Rather, we formulate everything in terms of the total free
energies of large collections of atoms. The only exception is the
case of a perfectly ordered and uniform system, for which we define
an atomic free energy as the total free energy divided by the number
of atoms.

\section{Background}

\label{sec:Background}

\subsection{The free energy of a harmonic solid}

The Gibbs energy $G_{N}$ of a single-component harmonic solid composed
of $N$ atoms is 
\begin{equation}
G_{N}=E_{N}+kT\sum_{i=1}^{3N}\ln\left(\dfrac{h\nu_{i}}{kT}\right),\label{eq:1}
\end{equation}
where $E_{N}$ is the static potential energy of the solid, $T$ is
temperature, $k$ is the Boltzmann factor, $h$ is Planck's constant,
and $\nu_{i}$ are normal vibrational frequencies. We omit the pressure
term, assuming that the pressure is zero for simplicity. Under this
assumption, the Gibbs free energy coincides with the Helmholtz free
energy and will be referred to as simply free energy.

If the solid is defect-free, then we can define its chemical potential
(free energy per atom) as $\mu_{0}=G_{N}/N$. We assume that the solid
is spatially uniform and subject to periodic boundary conditions,
and that it contains a large number of atoms ($N\gg1$). Accordingly,
the vibrational free energy accounted for by the second term in Eq.(\ref{eq:1})
can be approximated by an integral over the vibrational density of
states. Under these assumptions, $G_{N}$ is a homogeneous first-degree
function of $N$ and, therefore, $\mu_{0}$ does not depend on the
system size.

\subsection{Point-defect properties}

If a solid contains vacancies ($v$), their diffusion potential $M_{v}$
(relative to atoms) is given by 
\begin{equation}
M_{v}=\left[G_{N-1}^{(v)}-N\mu_{0}\right]+kT\ln c_{v}.\label{eq:2}
\end{equation}
Here, $c_{v}\ll1$ is the vacancy concentration defined as the fraction
of unoccupied lattice sites, and $G_{N-1}^{(v)}$ is the free energy
of a solid containing a single vacancy and $(N-1)$ atoms. The rigorous
definition of the diffusion potential was given by Larché and Cahn
\citep{Larche_Cahn_78,Larche73,Larche1985}. An application of this
concept to point defects was discussed in detail by Mullins and Sekerka
\citep{Mullins1985}.

The term in square brackets in Eq.(\ref{eq:2}) is the change in free
energy when an atom is removed from a perfect-lattice solid containing
$N$ atoms. The initial free energy of the solid is $N\mu_{0}$. It
is assumed that, after the atom is removed, the system is relaxed
to thermodynamic equilibrium by minimizing its total free energy.
If the system is sufficiently large ($N\gg1$), the term in the square
brackets is size-independent.

Vacancies can be formally treated as fictitious species with a chemical
potential $\mu_{v}$ defined as the free-energy change when a single
vacancy is created in a defect-free solid without changing the total
number of atoms. Thus, 
\begin{equation}
\mu_{v}=\left[G_{N-1}^{(v)}-(N-1)\mu_{0}\right]+kT\ln c_{v}.\label{eq:13}
\end{equation}
This formula does not require that the vacancies be in thermodynamic
equilibrium with respect to their creation-annihilation process. But
if they are in this type of equilibrium, then their concentration
$c_{v}^{\mathrm{eq}}$ is defined by the condition $\mu_{v}=0$. This
condition yields 
\begin{equation}
c_{v}^{\mathrm{eq}}=\exp\left(-\dfrac{\tilde{\mu}_{v}}{kT}\right),\label{eq:14}
\end{equation}
where 
\begin{equation}
\tilde{\mu}_{v}=G_{N-1}^{(v)}-(N-1)\mu_{0}\label{eq:15}
\end{equation}
is the non-configurational part of $\mu_{v}$, which is also called
the free energy of vacancy formation. To reach thermodynamic equilibrium,
the solid must be able to vary the number of lattice sites while keeping
a fixed number of atoms. This lattice site variation allows the solid
to minimize its total free energy by adjusting the number of vacancies.

Interstitial defects ($I$) are treated in a similar manner. We first
consider self-interstitial atoms. Their diffusion potential is not
well-defined, but their chemical potential is well-defined and given
by 
\begin{equation}
\mu_{I}=\left[G_{N+1}^{(I)}-(N+1)\mu_{0}\right]+kT\ln c_{I}.\label{eq:31}
\end{equation}
Here, $G_{N+1}^{(I)}$ is the free energy of a solid containing a
single self-interstitial atom. The interstitial concentration $c_{I}$
is defined as the occupied fraction of the available interstitial
sites. In contrast to the vacancy case, $G_{N+1}^{(I)}$ generally
includes not only the vibrational free energy but also an additional
term $-kT\ln\sigma$ associated with the orientational entropy $k\ln\sigma$.
The symmetry factor $\sigma$ represents the additional degeneracy
of the microstates due to different orientations of the interstitial
configuration. For example, a $\left\langle 100\right\rangle $ interstitial
dumbbell in FCC metals can have three energetically equivalent orientations
($\sigma=3$).

The treatment of interstitial impurity atoms is similar, except that
their concentration $c_{I}$ is an independent variable set by their
chemical potential outside the solid. Inside the solids, the chemical
potential of the impurity atoms is given by 
\begin{equation}
\mu_{I}=\left[G_{N}^{(I)}-N\mu_{0}\right]+kT\ln c_{I}.\label{eq:32}
\end{equation}
Here, $G_{N}^{(I)}$ is the free energy of a solid containing one
impurity atom and $N$ host atoms, and $\mu_{0}$ is the chemical
potential of the host atoms. As before, the free energy $G_{N}^{(I)}$
generally includes the orientational free energy. For example, impurity
atoms may form dumbbell configurations with host atoms, which can
have several energetically equivalent orientations.

For substitutional impurity atoms ($s$), both the diffusion potential
and the chemical potential are well-defined and given by the equations
\begin{equation}
M_{s}=\left[G_{N-1}^{(s)}-N\mu_{0}\right]+kT\ln c_{s},\label{eq:33}
\end{equation}
\begin{equation}
\mu_{s}=\left[G_{N-1}^{(s)}-(N-1)\mu_{0}\right]+kT\ln c_{s}.\label{eq:34}
\end{equation}
Here, $G_{N-1}^{(s)}$ is the free energy of a solid containing a
single substitutional impurity atom and $(N-1)$ host atoms. The impurity
concentration $c_{s}$ is the number of impurity atoms per lattice
site. This concentration is an independent variable controlled by
the chemical potential of the impurity atoms outside the solid.

\subsection{Transition-state theory}

\label{subsec:TST}

In transition-state theory (TST) \citep{Eyring:1935aa}, the rate
of transition (number of transitions per unit time) from one stable
or metastable state of a system to another is given by 
\begin{equation}
w=\dfrac{kT}{h}\exp\left(-\dfrac{g^{*}}{kT}\right),\label{eq:4}
\end{equation}
where $g^{*}$ is the activation free energy of the transition. Suppose
that the system is an $N$-atomic solid and the transition event is
localized, i.e., involves only a single atom or a small group of neighboring
atoms. Then 
\begin{equation}
g^{*}=G_{N}^{*}-G_{N},\label{eq:5}
\end{equation}
where $G_{N}$ is the free energy of the solid before the transition
and $G_{N}^{*}$ is its free energy in the transition state (activated
complex \citep{Eyring:1935aa}). For example, the initial state can
be a solid with a point defect undergoing an atomic rearrangement
such as a vacancy jump to a new position. As above, for a sufficiently
large solid ($N\gg1$), $g^{*}$ does not depend on $N$ and is a
local property.

In the harmonic TST \citep{Vineyard:1957vo}, the initial and activated
states are characterized by different energies $E_{N}$ and $E_{N}^{*}$,
different sets of vibrational frequencies $\nu_{i}$ and $\nu_{i}^{*}$,
and generally different symmetry factors $\sigma$ and $\sigma^{*}$.
The activated state has one vibrational mode less than the initial
state. This vibrational mode is replaced by translational motion through
the saddle point on the potential energy surface along the minimum-energy
path. Consequently, 
\begin{equation}
g^{*}=E_{N}^{*}-E_{N}+kT\ln\left({\displaystyle \dfrac{{\displaystyle \prod_{i=1}^{3N-1}\nu_{i}^{*}}}{{\displaystyle \prod_{i=1}^{3N}\nu_{i}}}}\dfrac{kT}{h}\dfrac{\sigma}{\sigma^{*}}\right).\label{eq:6}
\end{equation}
The transition rate becomes 
\begin{equation}
w=\nu_{0}\dfrac{\sigma^{*}}{\sigma}\exp\left(-\dfrac{\varepsilon^{*}}{kT}\right),\label{eq:7}
\end{equation}
where 
\begin{equation}
\varepsilon^{*}=E_{N}^{*}-E_{N}\label{eq:8}
\end{equation}
is the activation energy of the transition and 
\begin{equation}
\nu_{0}={\displaystyle \dfrac{{\displaystyle \prod_{i=1}^{3N}\nu_{i}}}{{\displaystyle \prod_{i=1}^{3N-1}\nu_{i}^{*}}}}\label{eq:9}
\end{equation}
is the attempt frequency.

For example, the rate of vacancy exchange with a neighboring atom
is given by Eq.(\ref{eq:4}) with 
\begin{equation}
g^{*}=G_{N-1}^{(v)*}-G_{N-1}^{(v)},\label{eq:10}
\end{equation}
where $G_{N-1}^{(v)}$ is the free energy of a solid that contains
$N$ sites with one vacancy, and $G_{N-1}^{(v)*}$ is the free energy
of the activated state of the vacancy jump. In the harmonic approximation,
the vacancy jump rate is given by Eq.(\ref{eq:7}) with the activation
energy 
\begin{equation}
\varepsilon^{*}=E_{N-1}^{(v)*}-E_{N-1}^{(v)}\label{eq:11}
\end{equation}
and attempt frequency 
\begin{equation}
\nu_{0}={\displaystyle \dfrac{{\displaystyle \prod_{i=1}^{3N-3}\nu_{i}}}{{\displaystyle \prod_{i=1}^{3N-4}\nu_{i}^{*}}}}.\label{eq:12}
\end{equation}
In most cases, vacancy jumps do not change orientational symmetry,
so $\sigma=\sigma^{*}$.

\subsection{Diffusion kinetics in solids}

\label{subsec:Diffusion-kinetics}

\subsubsection{Self-diffusion by the vacancy mechanism}

Let us first consider self-diffusion by the vacancy mechanism in a
single-component solid. Atoms diffuse by exchanges with vacancies.
An atom can make a diffusive jump only if there is a vacancy at a
neighboring lattice site and the atom exchanges positions with that
vacancy before the latter makes a dissociative jump and walks away.
The fraction of time in which a particular neighboring site is vacant
is $c_{v}$. The self-diffusion coefficient is given by \citep{Philibert,Mehrer2007}
\begin{equation}
D=f\xi l^{2}c_{v}w,\label{eq:16}
\end{equation}
where $f$ is the jump correlation factor, $\xi$ is a geometric factor,
$l$ is the vacancy jump length, and $w$ is the vacancy-atom exchange
rate. Both $f$ and $\xi$ are constants that depend only on the crystal
structure. The exchange rate $w$ is given by the TST equations (\ref{eq:4})
and (\ref{eq:5}), from which 
\begin{equation}
D=\dfrac{f\xi l^{2}kT}{h}c_{v}\exp\left(-\dfrac{G_{N-1}^{(v)*}-G_{N-1}^{(v)}}{kT}\right).\label{eq:17}
\end{equation}

Equation (\ref{eq:17}) does not require that the vacancies be in
thermodynamic equilibrium. But if they are, then we can use equations
(\ref{eq:14}) and (\ref{eq:15}) to obtain 
\begin{equation}
D=\dfrac{f\xi l^{2}kT}{h}\exp\left(-\dfrac{G_{N-1}^{(v)*}-(N-1)\mu_{0}}{kT}\right).\label{eq:18}
\end{equation}
In the harmonic approximation, 
\begin{equation}
D=f\xi l^{2}\nu_{0}\exp\left(-\dfrac{E_{N-1}^{(v)*}-(N-1)\varepsilon_{0}}{kT}\right).\label{eq:19}
\end{equation}
where $\varepsilon_{0}=E_{N}/N$ is the perfect-crystal energy per
atom. Importantly, the attempt frequency $\nu_{0}$ is given by an
equation similar to Eq.(\ref{eq:12}) except that $\nu_{i}$ are the
frequencies in a perfect lattice composed of $(N-1)$ atoms. Borisov
et al.~\citep{Borisov_1963} pointed out that, according to Eq.(\ref{eq:19}),
the diffusion coefficient depends only on the activated complex. It
does not depend on the vacancy formation energy or vacancy concentration,
as long as the latter corresponds to thermodynamic equilibrium with
respect to the vacancy creation and annihilation. 

In systems with non-equilibrium vacancy concentration, the latter
acts as an independent control parameter that influences the diffusion
coefficient. In this case, Eq.(\ref{eq:17}) remains valid, but Eq.(\ref{eq:19})
does not apply. In this work, we assume that the vacancy concentration
in the lattice is in thermodynamic equilibrium with the vacancy sinks
and sources, which allows us to use Eq.(\ref{eq:19}) and similar
equations throughout the paper. As discussed below, vacancies in GBs
are treated differently. They are assumed to be  in equilibrium with
the lattice but not in internal equilibrium with respect to site creation
and annihilation.

\subsubsection{Self-diffusion by indirect interstitial mechanisms}

Next, we consider self-diffusion by indirect interstitial mechanisms.
One of them is the interstitialcy mechanism \citep{Philibert,Mehrer2007},
in which an atom occupying an interstitial position displaces a neighboring
lattice atom from its regular position and takes its place. The displaced
atom becomes a self-interstitial atom and, in turn, displaces another
lattice atom, and the process continues. The atoms diffuse by jumping
from an interstitial position to a substitutional position and back
to an interstitial position. Each diffusive event is a simultaneous
displacement of two atoms.

Another case is the interstitial dumbbell mechanism. In some solid
materials, such as FCC metals, self-interstitials exist as $\left\langle 100\right\rangle $-oriented
dumbbells centered on a lattice site. The diffusive event is then
a simultaneous displacement of three atoms, resulting in a translation
of the center of the dumbbell to a neighboring lattice site. This
translation can be accompanied by a 90-degree rotation of the dumbbell.

In both the interstitialcy and dumbbell mechanisms, an atom spends
part of its time in a substitutional position and part in an interstitial
position. At each step of diffusion, a group of $n=2$ or $n=3$ atoms
undergoes a coordinated displacement along a straight line or follows
a nonlinear trajectory. Consequently, the activated complex is a linear
of nonlinear configuration of the $n$ atoms. The two types of atomic
rearrangements are referred to as, respectively, collinear and non-collinear
diffusive events. We will disregard the difference between them and
assume that the free energy of the activated complex is the same in
both cases.

Considering that the diffusion process is mediated by point defects
(interstitials), the diffusion coefficient is given by an expression
similar to the one for the vacancy mechanism: 
\begin{equation}
D=\dfrac{f\xi l^{2}kT}{h}c_{I}\exp\left(-\dfrac{G_{N+1}^{(I)*}-G_{N+1}^{(I)}}{kT}\right).\label{eq:22}
\end{equation}
Here, $G_{N+1}^{(I)}$ is the free energy of a solid containing $N$
lattice atoms and a single interstitial atom, $G_{N+1}^{(I)*}$ is
the free energy of this solid in the activated state of a diffusive
event, and $c_{I}$ is the interstitial concentration. If the latter
corresponds to thermodynamic equilibrium with respect to interstitial
creation and annihilation, then 
\begin{equation}
D=\dfrac{f\xi l^{2}kT}{h}\exp\left(-\dfrac{G_{N+1}^{(I)*}-(N+1)\mu_{0}}{kT}\right).\label{eq:23}
\end{equation}
In the harmonic approximation, 
\begin{equation}
D=f\xi l^{2}\nu_{0}\dfrac{\sigma^{*}}{\sigma}\exp\left(-\dfrac{E_{N+1}^{(I)*}-(N+1)\varepsilon_{0}}{kT}\right).\label{eq:23-1}
\end{equation}
As in the vacancy mechanism, the diffusion coefficient depends only
on the free energy of the activated complex.

\subsubsection{Impurity diffusion by the direct interstitial mechanism}

As another example, consider diffusion of impurity atoms by the direct
interstitial mechanism. The impurity concentration is fixed by the
chemical potential of the impurity atoms outside the solid. The impurity
atoms diffuse by hopping between interstitial positions. This mechanism
does not require any point defects other than the impurity atoms themselves.
Assuming that all interstitial sites are equivalent, the diffusion
coefficient is given by 
\begin{equation}
D=\xi l^{2}w=\dfrac{\xi l^{2}kT}{h}\exp\left(-\dfrac{G_{N}^{(I)*}-G_{N}^{(I)}}{kT}\right),\label{eq:20}
\end{equation}
where we applied the TST equation (\ref{eq:4}) to the hopping rate
$w$. In this equation, $G_{N}^{(I)}$ is the free energy of a solid
containing $N$ host atoms and one impurity atom, and $G_{N}^{(I)*}$
is the free energy when this atom is at the transition point between
two neighboring sites. Applying the harmonic TST, the diffusion coefficient
becomes 
\begin{equation}
D=\xi l^{2}\nu_{0}\exp\left(-\dfrac{E_{N}^{(I)*}-E_{N}^{(I)}}{kT}\right),\label{eq:21}
\end{equation}
where $E_{N}^{(I)}$ and $E_{N}^{(I)*}$ are the total energies before
the jump and in the activated state, respectively. The attempt frequency
is defined by Eq.(\ref{eq:9}) modified for the system of $(N+1)$
atoms ($N$ host atoms plus the impurity atom). We assumed that there
is no change in orientational symmetry during the transition ($\sigma^{*}=\sigma$).
Note the absence of the correlation factor in equations (\ref{eq:20})
and (\ref{eq:21}). This is because the sequential jumps of the impurity
atom are not correlated.

Formally, the above equations can be applied to self-diffusion by
direct hopping of self-interstitials. However, this diffusion mechanism
does not operate in real materials (unless in some exotic cases unknown
to us). Usually, self-interstitial atoms form dumbbells, crowdions,
or similar configurations involving both interstitial and lattice
atoms. Even when self-interstitial atoms are localized at interstitial
sites, they diffuse by indirect mechanisms involving lattice atoms.

As already mentioned, the above equations are based on the assumption
that all interstitial sites are equivalent. In reality, solids often
contain non-equivalent interstitial sites, such as octahedral and
tetrahedral. Impurity atoms hop between sites with different atomic
environments, a process that involves more than one activated complex.
Examples include impurity diffusion of hydrogen, carbon, and oxygen
in BCC and FCC metals. Such cases are not considered here.

\subsubsection{Impurity diffusion by the vacancy mechanism}

As a final example, consider diffusion of substitutional impurity
atoms by the vacancy mechanism. The diffusion coefficient is given
by \citep{Philibert,Mehrer2007} 
\begin{equation}
D=\dfrac{f\xi l^{2}kT}{h}p_{v}\exp\left(-\dfrac{G_{N-2}^{(vs)*}-G_{N-2}^{(vs)}}{kT}\right),\label{eq:25}
\end{equation}
where $G_{N-2}^{(vs)}$ and $G_{N-2}^{(vs)*}$ are the free energies
of a solid containing $(N-2)$ host atoms and a vacancy-impurity pair
($vs$) before the exchange of positions and in the activated state,
respectively. Eq.(\ref{eq:25}) looks similar to Eq.(\ref{eq:17})
for vacancy-mediated self-diffusion but differs from it in two ways.
Firstly, the correlation factor $f$ is no longer a geometric constant.
It depends on the vacancy-impurity interaction energy as a function
of the defect separation, as well as on temperature. Secondly, the
vacancy concentration $c_{v}$ is replaced by the vacancy availability
factor $p_{v}$ defined as the probability of finding a vacancy at
a given site next to the impurity atom. This factor can be approximated
by the expression 
\begin{equation}
p_{v}=c_{v}\exp\left(-\dfrac{g_{b}}{kT}\right),\label{eq:26}
\end{equation}
where $g_{b}$ is the vacancy-impurity binding free energy. The latter
is defined as the free energy difference between two configurations:
when the vacancy and impurity are nearest neighbors and when they
are far apart. Accordingly, 
\begin{equation}
g_{b}=G_{N-2}^{(vs)}-G_{N-1}^{(s)}-G_{N-1}^{(v)}+N\mu_{0},\label{eq:26-1}
\end{equation}
where $G_{N-1}^{(s)}$ and $G_{N-1}^{(v)}$ are the free energies
of solids with a single impurity atom and with a single vacancy, respectively.
As in previous cases, $g_{b}$ does not depend on $N$ when $N\gg1$.
If $g_{b}<0$, the impurity-vacancy interaction is attractive, leading
to an increased availability of vacancies next to the impurity atom
and, therefore, to accelerated diffusion. Thus, 
\begin{equation}
D=\dfrac{f\xi l^{2}kT}{h}c_{v}\exp\left(-\dfrac{G_{N-2}^{(vs)*}-G_{N-2}^{(vs)}+g_{b}}{kT}\right).\label{eq:27}
\end{equation}
This equation can be recast in a different form: 
\begin{equation}
D=\dfrac{f\xi l^{2}kT}{h}c_{v}\exp\left(-\dfrac{G_{N-2}^{(vs)*}-(N-2)\mu_{0}-[G_{N-1}^{(s)}-(N-1)\mu_{0}]-[G_{N-1}^{(v)}-(N-1)\mu_{0}]}{kT}\right).\label{eq:27-2}
\end{equation}
The two terms in the square brackets are the non-configurational parts
of the impurity and vacancy chemical potentials, respectively. For
example, the chemical potential of impurity atoms is 
\begin{equation}
\mu_{s}=G_{N-1}^{(s)}-(N-1)\mu_{0}+kT\ln c_{s},\label{eq:27-3}
\end{equation}
where $c_{s}$ is the impurity concentration (number per lattice site).

For the equilibrium vacancy concentration, the diffusion coefficient
becomes 
\begin{equation}
D=\dfrac{f\xi l^{2}kT}{h}\exp\left(-\dfrac{G_{N-2}^{(vs)*}-(N-2)\mu_{0}-[G_{N-1}^{(s)}-(N-1)\mu_{0}]}{kT}\right).\label{eq:27-1}
\end{equation}
In contrast to the self-diffusion case, the diffusion coefficient
depends not only on the free energy $G_{N-2}^{(vs)*}$ in the activated
state but also on the vacancy-impurity interaction through the correlation
factor.

\subsection{More on the activated complex}

\label{subsec:activated_complex}

\begin{figure}
\begin{centering}
\includegraphics[width=0.62\textwidth]{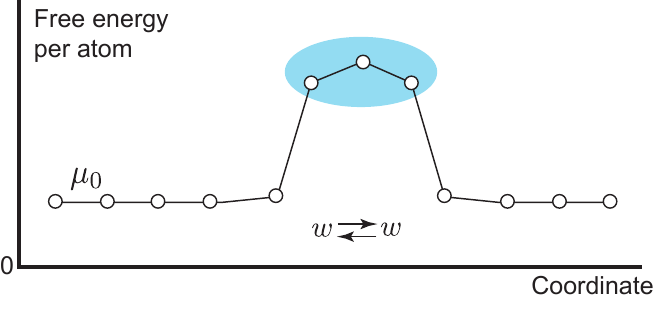} 
\par\end{centering}
\caption{Schematic representation of the atomic free energy as a function of
position in a crystalline solid with a point defect undergoing a transition
to a new state. The blue oval outlines the activated complex containing
$n=3$ atoms. The transition rate $w$ is given by equation (\ref{eq:132}).}
\label{fig:activated_complex} 
\end{figure}

Returning to the TST, we saw in the previous examples that the activated
complex can comprise more than one atom. Let the number of atoms in
the activated complex be $n$. For example, as mentioned above, $n=3$
for interstitial dumbbell jumps. The $n$ atoms that form the activated
complex are responsible for the higher free energy of the activated
state of the system than for a defect-free solid containing the same
number of atoms. This is shown schematically in Fig.~\ref{fig:activated_complex}
for $n=3$.

The subsequent discussions will involve the free energy, $\hat{g}$,
of the activated complex relative to the same reference state as for
$\mu_{0}$. Suppose a system composed of $N$ atoms contains a point
defect that jumps from one position to another through an activated
complex comprising $n$ atoms. Let the total free energy of the system
in the activated state be $G_{N}^{*}$. Then the free energy of the
activated complex is defined by 
\begin{equation}
\hat{g}=G_{N}^{*}-(N-n)\mu_{0},\label{eq:130}
\end{equation}
where $(N-n)$ is the number of atoms not affected by the activated
complex. Note that $\hat{g}$ does not depend on the total number
of atoms $N$ assuming $N\gg1$.

If $G_{N}$ is the total free energy of the system before the point-defect
jump, then the jump frequency is 
\begin{equation}
w=\dfrac{kT}{h}\exp\left(-\dfrac{G_{N}^{*}-G_{N}}{kT}\right),\label{eq:131}
\end{equation}
which can be rewritten in the form 
\begin{equation}
w=\dfrac{kT}{h}\exp\left(-\dfrac{\hat{g}}{kT}\right)\exp\left(\dfrac{G_{N}-(N-n)\mu_{0}}{kT}\right).\label{eq:132}
\end{equation}
This important equation will be used in the subsequent analysis.

Note that the value of $\hat{g}$ can be subjective because the activated
complex is often surrounded by lattice atoms that are significantly
displaced from their equilibrium positions and also contribute to
the excess free energy. In some cases, a judgment call has to be made
on whether or not to include an atom, or a group of atoms, in the
definition of the activated complex. This choice affects the number
of activated atoms $n$ and thus their free energy $\hat{g}$.

\section{Relationships between the grain boundary free energy and diffusion}

\label{sec:Borisov}

\subsection{The grain boundary model}

We will apply the theory reviewed in the previous section to GB diffusion.
The GB is represented by a slab composed of a \emph{fixed} number
$m$ of parallel atomic planes, each with a \emph{fixed} number of
sites. The atomic structure of the GB is assumed to be uniform but
generally different from the crystalline structure in the grains.
However, we assume that the GB structure supports the same point defects
as the crystalline lattice, including vacancies and interstitials.
We further assume that GB diffusion occurs by the same mechanism as
lattice diffusion, although the parameters of this mechanism can differ.

An important difference between the GB and the surrounding lattice
is that the number of GB sites cannot change, whereas the number of
lattice sites is allowed to vary by processes such as the generation
and annihilation of vacancies and self-interstitials (e.g., at the
surface of the sample). One consequence of this distinction is that
the grains can reach an equilibrium concentration of vacancies and
self-interstitials. In contrast, the concept of intrinsic point-defect
equilibrium in the GB has no meaning since the GB sites cannot be
created or destroyed. Nevertheless, we can talk about thermodynamic
equilibrium \emph{between} the GB and the grains, regardless of whether
the point defects in the grains are at equilibrium or not. The GB-lattice
equilibrium is expressed by the equality of the diffusion potentials
of substitutional point defects (e.g., vacancies or substitutional
impurity atoms) and the equality of chemical potentials of interstitial
point defects (e.g., self-interstitial or interstitial impurity atoms).

\subsection{Self-diffusion by the vacancy mechanism}

\label{subsec:Self-diffusion-vacancy}

The vacancy equilibrium between the GB and the lattice is expressed
by the equality of the diffusion potentials of vacancies: $M_{v}^{\prime}=M_{v}$.
Here and everywhere below, the prime superscript indicates GB properties.
Using Eq.(\ref{eq:2}), we can write 
\begin{equation}
\left[G_{N^{\prime}-1}^{(v)\prime}-N^{\prime}\mu_{0}^{\prime}\right]+kT\ln c_{v}^{\prime}=\left[G_{N-1}^{(v)}-N\mu_{0}\right]+kT\ln c_{v}.\label{eq:35}
\end{equation}
Here, $c_{v}^{\prime}$ and $c_{v}$ are the vacancy concentrations
in the GB and in the lattice, respectively. Recall that we assume
$N^{\prime}\gg1$ and $N\gg1$, so that the terms in square brackets
represent local properties that do not depend on the number of atoms
in the GB ($N^{\prime}$) and in the lattice ($N$). Eq.(\ref{eq:35})
allows us to predict the equilibrium ratio $c_{v}^{\prime}/c_{v}$:
\begin{equation}
\dfrac{c_{v}^{\prime}}{c_{v}}=\exp\left(-\dfrac{\left[G_{N^{\prime}-1}^{(v)\prime}-N^{\prime}\mu_{0}^{\prime}\right]-\left[G_{N-1}^{(v)}-N\mu_{0}\right]}{kT}\right).\label{eq:38}
\end{equation}

The lattice self-diffusion coefficient by the vacancy mechanism is
given by Eq.(\ref{eq:17}), which is repeated below for convenience:
\begin{equation}
D=\dfrac{f\xi l^{2}kT}{h}c_{v}\exp\left(-\dfrac{G_{N-1}^{(v)*}-G_{N-1}^{(v)}}{kT}\right).\label{eq:40}
\end{equation}
The GB diffusion coefficient is given by a similar equation: 
\begin{equation}
D^{\prime}=\dfrac{f^{\prime}\xi^{\prime}l^{\prime2}kT}{h}c_{v}^{\prime}\exp\left(-\dfrac{G_{N^{\prime}-1}^{(v)\prime*}-G_{N^{\prime}-1}^{(v)\prime}}{kT}\right).\label{eq:39}
\end{equation}
The ratio of the two diffusion coefficients is 
\begin{equation}
\dfrac{D^{\prime}}{D}=\dfrac{f^{\prime}\xi^{\prime}l^{\prime2}}{f\xi l^{2}}\left(\dfrac{c_{v}^{\prime}}{c_{v}}\right)\exp\left(-\dfrac{\left[G_{N^{\prime}-1}^{(v)\prime*}-G_{N^{\prime}-1}^{(v)\prime}\right]-\left[G_{N-1}^{(v)*}-G_{N-1}^{(v)}\right]}{kT}\right).\label{eq:41}
\end{equation}
Inserting $c_{v}^{\prime}/c_{v}$ from Eq.(\ref{eq:38}), we have
\begin{equation}
\dfrac{D^{\prime}}{D}=\dfrac{f^{\prime}\xi^{\prime}l^{\prime2}}{f\xi l^{2}}\exp\left(-\dfrac{\left[G_{N^{\prime}-1}^{(v)\prime*}-(N^{\prime}-1)\mu_{0}^{\prime}\right]-\left[G_{N-1}^{(v)*}-(N-1)\mu_{0}\right]-(\mu_{0}^{\prime}-\mu_{0})}{kT}\right).\label{eq:42}
\end{equation}

\begin{figure}
\begin{centering}
\includegraphics[width=0.92\textwidth]{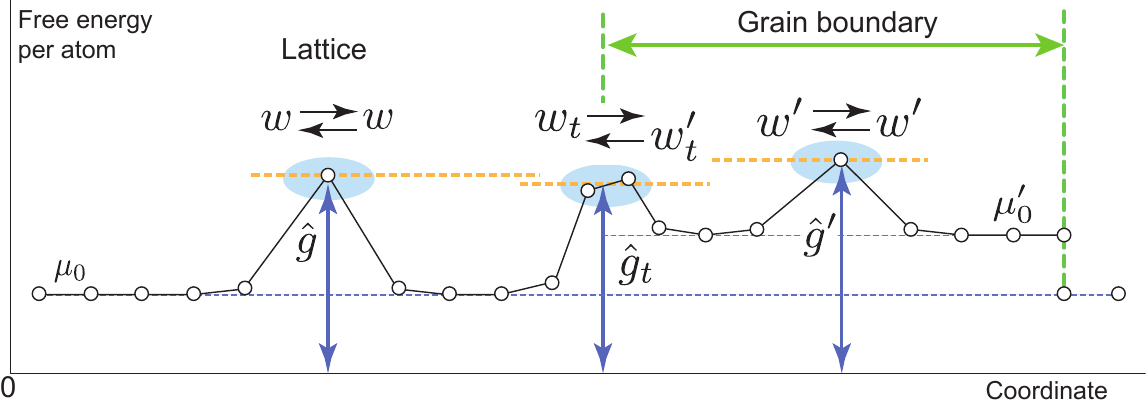}
\par\end{centering}
\caption{Schematic representation of free energies of the activated complexes
for vacancy jumps in the lattice ($\hat{g}$), in the GB ($\hat{g}^{\prime}$),
and between the GB and the lattice ($\hat{g}_{t}$). The plot shows
the free energies of atoms as a function of position in the perfect
lattice ($\mu_{0}$), in the GB ($\mu_{0}^{\prime}$), and in the
activated complexes. The atomic jump rates are $w$ in the lattice,
$w^{\prime}$ in the GB, and $w_{t}$ and $w_{t}^{\prime}$ into and
out of the GB, respectively. All free energies are defined relative
to the same reference state. The blue ovals symbolize the activated
complexes. The orange dashed lines demonstrate that $\hat{g}$, $\hat{g}^{\prime}$,
and $\hat{g}_{t}$ could be close to each other but are generally
different. An activated complex may contain a single atom or a group
of several atoms undergoing a collective rearrangement. The diagram
is purely conceptual. In reality, free energies of individual atoms
are not well-defined, which is why the equations in the main text
are formulated in terms of total free energies of regions containing
many atoms.}
\label{fig:1}
\end{figure}

Next, we express $D^{\prime}/D$ through the free energies of the
activated complexes, $\hat{g}$ and $\hat{g}^{\prime}$, in the lattice
and in the GB relative to the same reference state. The choice of
the reference state is arbitrary and is dictated by convenience. For
example, we can choose a system of isolated atoms as a reference state.
The respective free energies of the activated complexes can be obtained
from Eq.(\ref{eq:130}), which gives 
\begin{equation}
\hat{g}=G_{N-1}^{(v)*}-(N-1-n)\mu_{0},\label{eq:36}
\end{equation}
\begin{equation}
\hat{g}^{\prime}=G_{N^{\prime}-1}^{(v)\prime*}-(N^{\prime}-1-n)\mu_{0}^{\prime}.\label{eq:37}
\end{equation}
As before, $\hat{g}$ and $\hat{g}^{\prime}$ are independent of the
numbers of atoms $N$ and $N^{\prime}$ in the limits of $N\gg1$
and $N^{\prime}\gg1$. We have assumed that the activated complex
contains the same number of atoms $n$ in the GB and in the lattice.
The diagram in Fig.~\ref{fig:1} gives a graphical representation
of $\hat{g}$ and $\hat{g}^{\prime}$. (The jump rates $w$, $w^{\prime}$,
$w_{t}$, and $w_{t}^{\prime}$ shown on this diagram are discussed
in the Appendix.)

From equations (\ref{eq:36}) and (\ref{eq:37}), we have 
\begin{equation}
\left[G_{N^{\prime}-1}^{(v)\prime*}-(N^{\prime}-1)\mu_{0}^{\prime}\right]-\left[G_{N-1}^{(v)*}-(N-1)\mu_{0}\right]=(\hat{g}^{\prime}-\hat{g})-n(\mu_{0}^{\prime}-\mu_{0}).\label{eq:43}
\end{equation}
Inserting this expression into Eq.(\ref{eq:42}), we obtain the important
equation: 
\begin{equation}
\dfrac{D^{\prime}}{D}=\left[\dfrac{f^{\prime}\xi^{\prime}l^{\prime2}}{f\xi l^{2}}\exp\left(-\dfrac{\hat{g}^{\prime}-\hat{g}}{kT}\right)\right]\exp\left(\dfrac{(n+1)(\mu_{0}^{\prime}-\mu_{0})}{kT}\right).\label{eq:44}
\end{equation}
This equation could also be derived from Eq.(\ref{eq:132}).

Equation (\ref{eq:44}) is the starting point for several approximations
eventually leading to the Borisov relation. First of all, the pre-exponential
factor $(f^{\prime}\xi^{\prime}l^{\prime2}/f\xi l^{2})$ is less important
than the exponential factors and can be dropped. Then the ratio $D^{\prime}/D$
depends on three properties: (1) the free energy difference $(\hat{g}^{\prime}-\hat{g})$
between the activated complexes of the vacancy jumps in the GB and
in the lattice, (2) the size of the activated complex ($n$), and
(3) the difference $(\mu_{0}^{\prime}-\mu_{0})$ between the chemical
potentials of the atoms in the GB and in the lattice. The latter difference
is not related to diffusion; it does not involve point defects and
is a purely thermodynamic property of the GB.

Borisov et al.~\citep{Borisov} implicitly assumed that $n=1$, which
is reasonable for the vacancy mechanism. Their next approximation
was to consider the free energies $\hat{g}^{\prime}$ and $\hat{g}$
equal or at least close enough to approximate the exponential term
in the square brackets by unity. In other words, they assumed that
the activated complex of vacancy jumps does not depend (or does not
depend much) on whether the jumps occur in the GB or in the perfect
lattice. This is a very strong approximation that requires validation
by experiments and/or simulations. It will be discussed in more detail
later. For now, we will accept it and rewrite Eq.(\ref{eq:44}) in
the form 
\begin{equation}
\dfrac{D^{\prime}}{D}=\Lambda_{g}\exp\left(\dfrac{2(\mu_{0}^{\prime}-\mu_{0})}{kT}\right),\label{eq:45}
\end{equation}
where $\Lambda_{g}$ is a constant (close to unity) representing the
square brackets in Eq.(\ref{eq:44}).

Since $(\mu_{0}^{\prime}-\mu_{0})$ is the excess free energy of the
GB atoms relative to the lattice atoms, it is related to the GB free
energy $\gamma$ (per unit area) by 
\begin{equation}
\gamma=\dfrac{(\mu_{0}^{\prime}-\mu_{0})\delta}{\Omega^{\prime}},\label{eq:46}
\end{equation}
where $\Omega^{\prime}$ is the atomic volume in the GB. Then Eq.(\ref{eq:45})
becomes 
\begin{equation}
\dfrac{D^{\prime}}{D}=\Lambda_{g}\exp\left(\dfrac{2\gamma\Omega^{\prime}}{\delta kT}\right).\label{eq:47}
\end{equation}

Equation (\ref{eq:47}) links the normalized GB diffusivity to the
GB free energy. Since the exponent in this equation is positive, $D^{\prime}$
is greater than $D$. The higher the GB free energy, the faster the
GB diffusion. Qualitatively, this equation suggests that GB diffusion
is fast because the excess GB energy pushes the vacancies closer to
their activated state, reducing the activation barrier for their jumps.
Eq.(\ref{eq:47}) also suggests that fast GB diffusion does not require
that the free energy of vacancy formation in the GB be lower or higher
than in the lattice. According to this equation, the free energy of
vacancy formation does not affect the GB diffusivity. Only the transition
state matters.

Equation (\ref{eq:47}) can be expressed through the excess GB energy
$u$ rather than GB free energy $\gamma$: 
\begin{equation}
u=\dfrac{(\varepsilon_{0}^{\prime}-\varepsilon_{0})\delta}{\Omega^{\prime}},\label{eq:48}
\end{equation}
The equation for $D^{\prime}/D$ becomes 
\begin{equation}
\dfrac{D^{\prime}}{D}=\Lambda_{\varepsilon}\exp\left(\dfrac{2u\Omega^{\prime}}{\delta kT}\right),\label{eq:49}
\end{equation}
where the coefficient $\Lambda_{\varepsilon}$ absorbs the vibrational
entropies of the GB and lattice atoms. Like $\Lambda_{g}$, this coefficient
is treated as a constant.

\subsection{Impurity diffusion by the vacancy mechanism}

\label{subsec:Impurity-diffusion}

Next, we consider diffusion of substitutional impurity atoms by the
vacancy mechanism. The diffusion potential of the impurity atoms is
given by Eq.(\ref{eq:33}). From the GB-lattice equilibrium condition
$M_{s}^{\prime}=M_{s}$, we obtain the linear segregation isotherm
\begin{equation}
\dfrac{c_{s}^{\prime}}{c_{s}}=\exp\left(-\dfrac{g_{s}}{kT}\right),\label{eq:62}
\end{equation}
where

\begin{equation}
g_{s}=\left[G_{N^{\prime}-1}^{(s)\prime}-N^{\prime}\mu_{0}^{\prime}\right]-\left[G_{N-1}^{(s)}-N\mu_{0}\right]\label{eq:63}
\end{equation}
is the segregation free energy.

The diffusion mechanism involves swapping positions of impurity atoms
with vacancies. The activated complex is a nearest-neighbor vacancy-impurity
pair ($vs$) in the transition state. The free energy of this activated
complex is

\begin{equation}
\hat{g}=\left[G_{N-2}^{(vs)*}-(N-1-n)\mu_{0}\right]\label{eq:64}
\end{equation}
in the lattice and 
\begin{equation}
\hat{g}^{\prime}=\left[G_{N^{\prime}-2}^{(vs)\prime*}-(N^{\prime}-1-n)\mu_{0}^{\prime}\right]\label{eq:65}
\end{equation}
in the GB. In these equations, $G_{N-2}^{(vs)*}$ and $G_{N^{\prime}-2}^{(vs)\prime*}$
are the total free energies of the system in which a vacancy-impurity
pair is in the transition state in the lattice and in the GB, respectively.
To bring the system to the activated state starting from a perfect
crystal, one host atom is removed to create a vacancy and another
is removed to replace it with an impurity atom; hence the indices
$(N-2)$ and $(N^{\prime}-2)$. $n$ is the number of atoms in the
activated complex, including the impurity atom.

By algebraic rearrangements, the above equations can be rewritten
in the form 
\begin{equation}
\hat{g}=\left[G_{N-2}^{(vs)*}-G_{N-2}^{(vs)}\right]+\left[G_{N-1}^{(s)}-N\mu_{0}\right]+\left[G_{N-1}^{(v)}-N\mu_{0}\right]+g_{b}+(n+1)\mu_{0},\label{eq:66-1}
\end{equation}
\begin{equation}
\hat{g}^{\prime}=\left[G_{N^{\prime}-2}^{(vs)\prime*}-G_{N^{\prime}-2}^{(vs)\prime}\right]+\left[G_{N^{\prime}-1}^{(s)\prime}-N^{\prime}\mu_{0}^{\prime}\right]+\left[G_{N^{\prime}-1}^{(v)\prime}-N^{\prime}\mu_{0}^{\prime}\right]+g_{b}^{\prime}+(n+1)\mu_{0}^{\prime},\label{eq:67}
\end{equation}
where $g_{b}$ and $g_{b}^{\prime}$ are the vacancy-impurity binding
free energies given by 
\begin{equation}
g_{b}=G_{N-2}^{(vs)}-G_{N-1}^{(s)}-G_{N-1}^{(v)}+N\mu_{0},\label{eq:66}
\end{equation}
\begin{equation}
g_{b}^{\prime}=G_{N^{\prime}-2}^{(vs)\prime}-G_{N^{\prime}-1}^{(s)\prime}-G_{N^{\prime}-1}^{(v)\prime}+N^{\prime}\mu_{0}^{\prime}.\label{eq:67-2}
\end{equation}
Here, $G_{N-1}^{(vs)}$ and $G_{N^{\prime}-1}^{(vs)\prime}$ are the
respective free energies before the transition. Subtracting Eq.(\ref{eq:66-1})
from Eq.(\ref{eq:67}), we have 
\begin{eqnarray}
\hat{g}^{\prime}-\hat{g} & = & \left[G_{N^{\prime}-2}^{(vs)\prime*}-G_{N^{\prime}-2}^{(vs)\prime}\right]-\left[G_{N-2}^{(vs)*}-G_{N-2}^{(vs)}\right]\nonumber \\
 & + & \left[G_{N^{\prime}-1}^{(v)\prime}-N^{\prime}\mu_{0}^{\prime}\right]-\left[G_{N-1}^{(v)}-N\mu_{0}\right]\nonumber \\
 & + & g_{s}+(g_{b}^{\prime}-g_{b})+(n+1)(\mu_{0}^{\prime}-\mu_{0}).\label{eq:75}
\end{eqnarray}

In section \ref{subsec:Diffusion-kinetics}, we found that the impurity
diffusion coefficient is given by Eq.(\ref{eq:27}). Applying that
equation, we obtain the diffusivity ratio: 
\begin{equation}
\dfrac{D^{\prime}}{D}=\dfrac{f^{\prime}\xi^{\prime}l^{\prime2}}{f\xi l^{2}}\left(\dfrac{c_{v}^{\prime}}{c_{v}}\right)\exp\left(-\dfrac{\left[G_{N^{\prime}-2}^{(vs)\prime*}-G_{N^{\prime}-2}^{(vs)\prime}\right]-\left[G_{N-2}^{(vs)*}-G_{N-2}^{(vs)}\right]+(g_{b}^{\prime}-g_{b})}{kT}\right).\label{eq:68}
\end{equation}
Inserting the vacancy concentration ratio from Eq.(\ref{eq:38}),
we have 
\begin{eqnarray}
\dfrac{D^{\prime}}{D} & = & \dfrac{f^{\prime}\xi^{\prime}l^{\prime2}}{f\xi l^{2}}\exp\left(-\dfrac{\left[G_{N^{\prime}-2}^{(vs)\prime*}-G_{N^{\prime}-2}^{(vs)\prime}\right]-\left[G_{N-2}^{(vs)*}-G_{N-2}^{(vs)}\right]}{kT}\right)\nonumber \\
 & \times & \exp\left(-\dfrac{\left[G_{N^{\prime}-1}^{(v)\prime}-\mu_{0}^{\prime}N^{\prime}\right]-\left[G_{N-1}^{(v)}-\mu_{0}N\right]+(g_{b}^{\prime}-g_{b})}{kT}\right).\label{eq:69}
\end{eqnarray}
Combining this equation with Eq.(\ref{eq:75}), we finally obtain

\begin{equation}
\dfrac{D^{\prime}}{D}=\left[\dfrac{f^{\prime}\xi^{\prime}l^{\prime2}}{f\xi l^{2}}\exp\left(-\dfrac{\hat{g}^{\prime}-\hat{g}}{kT}\right)\right]\exp\left(\dfrac{g_{s}+(n+1)(\mu_{0}^{\prime}-\mu_{0})}{kT}\right).\label{eq:70}
\end{equation}

Equation (\ref{eq:70}) looks similar to Eq.(\ref{eq:44}) for self-diffusion
by the vacancy mechanism but includes the addition term $g_{s}$ representing
the segregation free energy. In the case of self-diffusion, the impurity
atoms are chemically identical to the host atoms and $g_{s}$ is zero.
In this case, Eq.(\ref{eq:70}) correctly reduces to Eq.(\ref{eq:44}).

As in the previous cases, we can assume that the expression in the
square brackets in Eq.(\ref{eq:70}) is constant. Then Eq.(\ref{eq:70})
then takes the approximate form 
\begin{equation}
\dfrac{D^{\prime}}{D}=\Lambda_{g}\exp\left(\dfrac{g_{s}+(n+1)(\mu_{0}^{\prime}-\mu_{0})}{kT}\right)=\Lambda_{g}\exp\left(\dfrac{(n+1)\gamma\Omega^{\prime}/\delta+g_{s}}{kT}\right).\label{eq:71}
\end{equation}
The assumption of constant $\Lambda_{g}$ is more risky than before
because the correlation factors are no longer geometric constants.
They can vary significantly with temperature and be very different
in the GB and inside the grains. Furthermore, although the vacancy-impurity
binding energies do not appear in Eq.(\ref{eq:70}) explicitly, they
are included in the transition-state free energies $g^{\prime}$ and
$g$. Since they appear in the exponent, their impact on the diffusion
coefficients through $g^{\prime}$ and $g$ can be significant.

\subsection{Impurity diffusion by the direct interstitial mechanism}

\label{subsec:direct-interst}

Let us next consider diffusion of dilute impurity atoms by the direct
interstitial mechanism. The impurity atoms diffuse by hopping between
energetically equivalent interstitial sites. The chemical potentials
of the impurity atoms in the lattice and in the GB are, respectively,

\begin{equation}
\mu_{I}=\left[G_{N}^{(I)}-N\mu_{0}\right]+kT\ln c_{I},\label{eq:50}
\end{equation}
\begin{equation}
\mu_{I}^{\prime}=\left[G_{N\prime}^{(I)\prime}-N^{\prime}\mu_{0}^{\prime}\right]+kT\ln c_{I}^{\prime}.\label{eq:51}
\end{equation}
The terms in square brackets are local properties (independent of
$N$ and $N^{\prime}$ when $N\gg1$ and $N^{\prime}\gg1$) and have
the meaning of the solubility free energies of the impurity atoms
in the lattice and in the GB, respectively. The equilibrium distribution
of the impurity atoms between the GB and the lattice is reached when
$\mu_{I}^{\prime}=\mu_{I}$. This condition yields the linear GB segregation
isotherm 
\begin{equation}
\dfrac{c_{I}^{\prime}}{c_{I}}=\exp\left(-\dfrac{g_{s}}{kT}\right),\label{eq:52}
\end{equation}
where

\begin{equation}
g_{s}=\left[G_{N\prime}^{(I)\prime}-N^{\prime}\mu_{0}^{\prime}\right]-\left[G_{N}^{(I)}-N\mu_{0}\right]\label{eq:52-1}
\end{equation}
is the segregation free energy. For a segregating impurity, $g_{s}<0$
and thus $c_{I}^{\prime}>c_{I}$.

\begin{figure}
\begin{centering}
\includegraphics[width=0.93\textwidth]{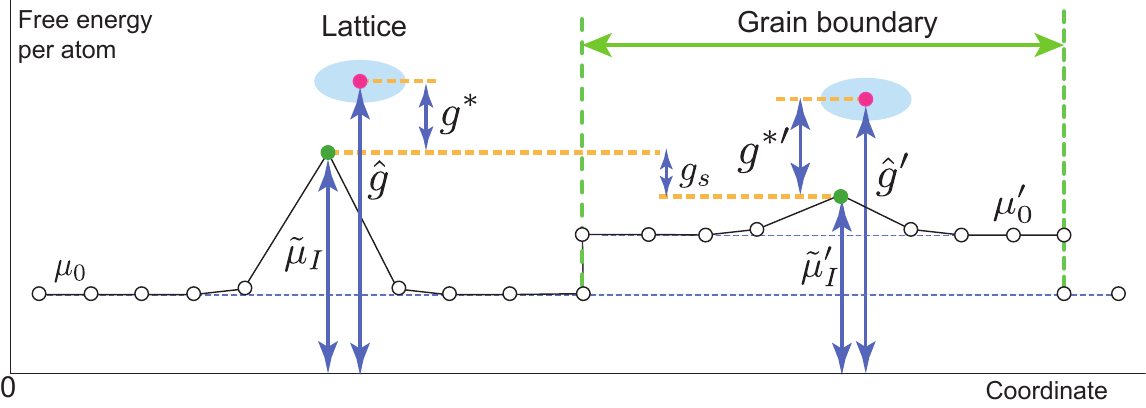}
\par\end{centering}
\caption{Schematic representation of the free energies of interstitial impurity
atoms during diffusion by the direct interstitial mechanism. The plot
shows the free energies of atoms as a function of position. The host
atoms are represented by white circles, and the interstitial atoms
before their jumps and in the activated state are represented by green
and red circles, respectively. The blue ovals symbolize the activated
complexes, whose free energies are $\hat{g}$ and $\hat{g}^{\prime}$.
$\tilde{\mu}_{I}$ and $\tilde{\mu}_{I}^{\prime}$ are the non-configurational
parts of the chemical potentials of the impurity atoms. $g^{*}$ and
$g^{*\prime}$ are the activation barriers of the interstitial jumps.
Note that $\tilde{\mu}_{I}^{\prime}<\tilde{\mu}_{I}$, leading to
the negative segregation energy $g_{s}=\tilde{\mu}_{I}^{\prime}-\tilde{\mu}_{I}<0$
and impurity segregation at the GB. Assuming that $\hat{g}^{\prime}\approx\hat{g}$,
the activation free energy of GB diffusion is higher than that of
lattice diffusion: $g^{*\prime}>g^{*}$. The diagram is purely conceptual.
In reality, free energies of individual atoms are not well-defined.}
\label{fig:2}
\end{figure}

Next, we define the free energies of the activated complexes relative
to a common reference state (Fig.~\ref{fig:2}):

\begin{equation}
\hat{g}=\left[G_{N}^{(I)*}-(N+1-n)\mu_{0}\right],\label{eq:53}
\end{equation}
\begin{equation}
\hat{g}^{\prime}=\left[G_{N^{\prime}}^{(I)\prime*}-(N^{\prime}+1-n)\mu_{0}^{\prime}\right].\label{eq:54}
\end{equation}
Here, $n$ is the number of atoms in the activated complex, including
the impurity atom. As before, $n$ is assumed to be the same in the
GB and in the lattice. Equations (\ref{eq:53}) and (\ref{eq:54})
can be rewritten in the form: 
\begin{equation}
\hat{g}=\left[G_{N}^{(I)*}-G_{N}^{(I)}\right]+\left[G_{N}^{(I)}-N\mu_{0}\right]+(n-1)\mu_{0},\label{eq:55}
\end{equation}
\begin{equation}
\hat{g}^{\prime}=\left[G_{N^{\prime}}^{(I)\prime*}-G_{N^{\prime}}^{(I)\prime}\right]+\left[G_{N^{\prime}}^{(I)\prime}-N^{\prime}\mu_{0}^{\prime}\right]+(n-1)\mu_{0}^{\prime}.\label{eq:56}
\end{equation}

In the next step, we use Eq.(\ref{eq:20}) for the impurity diffusion
coefficient to obtain 
\begin{equation}
\dfrac{D^{\prime}}{D}=\dfrac{\xi^{\prime}l^{\prime2}}{\xi l^{2}}\exp\left(-\dfrac{\left[G_{N^{\prime}}^{(I)\prime*}-G_{N^{\prime}}^{(I)\prime}\right]-\left[G_{N}^{(I)*}-G_{N}^{(I)}\right]}{kT}\right).\label{eq:57}
\end{equation}
Combining this equation with equations (\ref{eq:52-1}), (\ref{eq:55})
and (\ref{eq:56}), we arrive at 
\begin{equation}
\dfrac{D^{\prime}}{D}=\left[\dfrac{\xi^{\prime}l^{\prime2}}{\xi l^{2}}\exp\left(-\dfrac{\hat{g}^{\prime}-\hat{g}}{kT}\right)\right]\exp\left(\dfrac{g_{s}}{kT}\right)\exp\left(\dfrac{(n-1)(\mu_{0}^{\prime}-\mu_{0})}{kT}\right).\label{eq:58}
\end{equation}
As for the vacancy mechanism, the pre-exponential factor in square
brackets is assumed to be a constant $\Lambda_{g}$. Expressing the
free energy difference $(\mu_{0}^{\prime}-\mu_{0})$ through the GB
free energy $\gamma$, we finally obtain: 
\begin{equation}
\dfrac{D^{\prime}}{D}=\Lambda_{g}\exp\left(\dfrac{(n-1)\gamma\Omega^{\prime}/\delta+g_{s}}{kT}\right).\label{eq:59}
\end{equation}
Alternatively, Eq.(\ref{eq:59}) can be rewritten in the form 
\begin{equation}
\dfrac{D^{\prime}}{D}=\Lambda_{g}\dfrac{c_{I}}{c_{I}^{\prime}}\exp\left(\dfrac{(n-1)\gamma\Omega^{\prime}}{\delta kT}\right).\label{eq:60}
\end{equation}

Equation (\ref{eq:59}) is similar to Eq.(\ref{eq:71}) for impurity
diffusion by the vacancy mechanism, except that the factor in front
of $\gamma$ is now $(n-1)$ instead of $(n+1)$. Since the mechanism
is the hopping of single impurity atoms, $n=1$ is a reasonable approximation
and 
\begin{equation}
\dfrac{D^{\prime}}{D}=\Lambda_{g}\exp\left(\dfrac{g_{s}}{kT}\right).\label{eq:59-1}
\end{equation}
Note that $D^{\prime}/D$ does not depend on the GB free energy. It
can no longer be used to extract $\gamma$ from diffusion data. Furthermore,
Eq.(\ref{eq:59-1}) predicts that in the presence of GB segregation
($g_{s}<0$), GB diffusion is \emph{slower} than lattice diffusion.
This is a manifestation of the GB trapping effect. Qualitatively,
the impurity atoms have a free energy in the GB lower than that in
the lattice, which separates them further from the transition state.

\subsection{Self-diffusion by indirect interstitial mechanisms}

Finally, consider self-diffusion by the interstitialcy or dumbbell
interstitial mechanisms. In both cases, the interstitial atoms are
chemically identical to the host atoms. Their concentrations are still
given by Eq.(\ref{eq:52}), where $g_{s}$ is the free energy difference
between the self-interstitial positions in the GB and in the lattice.
However, we should now consider self-diffusion of \emph{all} atoms
regardless of their occupation. The diffusion process is mediated
by point defects (self-interstitials), but the diffusing atoms visit
both interstitial and substitutional sites. As discussed in section
\ref{subsec:Diffusion-kinetics}, the diffusion coefficient is then
proportional to the interstitial concentration $c_{I}$, as indicated
in Eq.(\ref{eq:22}). Therefore, the diffusivity ratio $D^{\prime}/D$
in Eq.(\ref{eq:60}) must be multiplied by the factor $c_{I}^{\prime}/c_{I}$.
This factor cancels with $c_{I}^ {}/c_{I}^{\prime}$, which yields
\begin{equation}
\dfrac{D^{\prime}}{D}=\Lambda_{g}\exp\left(\dfrac{(n-1)\gamma\Omega^{\prime}}{\delta kT}\right).\label{eq:61}
\end{equation}
The appropriate values of $n$ are $n=2$ for the interstitialcy mechanism
and $n=3$ for the dumbbell mechanism.

\section{Discussion}

\label{sec:Discussion}

\subsection{The significance of the derived equations}

We have presented a detailed, step-by-step derivation of a relationship
between the GB diffusion coefficient and the GB free energy. The relationship
depends on the diffusion mechanism and is different for self-diffusion
and diffusion of solute atoms. In analyzing the alloy case, we only
considered diffusion of impurity atoms with a small concentration.

The general form of the relationship is 
\begin{equation}
\dfrac{D^{\prime}}{D}=\Lambda_{g}\exp\left(\dfrac{g_{s}+\alpha(\mu_{0}^{\prime}-\mu_{0})}{kT}\right).\label{eq:72}
\end{equation}
Here, $\alpha=n-1$ for the interstitial diffusion mechanism, $\alpha=n+1$
for the vacancy diffusion mechanism, $g_{s}$ is the free energy of
GB segregation and $(\mu_{0}^{\prime}-\mu_{0})$ is the chemical potential
difference between the GB atoms and the lattice atoms in a single-component
solid (solvent in alloys). For self-diffusion in elemental solids,
$g_{s}=0$. The number of atoms in the activated complex ($n$) depends
on details of the diffusion mechanism. For example, $n=1$ is a reasonable
assumption for the vacancy mechanism and for diffusion of impurity
atoms by direct hopping between interstitial sites. For indirect interstitial
mechanisms, $n>1$ is more appropriate.

The pre-exponential factor $\Lambda_{g}$ in Eq.(\ref{eq:72}) depends
on the vibrational frequencies in the defect-free and activated states
of the system, the jump correlation factors, and the atomic structures
of the GB and of the lattice. In practical applications, this factor
is assumed to be constant. In most cases, it is assumed to be unity
due to a lack of better estimates. By splitting the free energies
into energy and entropy terms, Eq.(\ref{eq:72}) can be rewritten
in the energy form: 
\begin{equation}
\dfrac{D^{\prime}}{D}=\Lambda_{\varepsilon}\exp\left(\dfrac{\varepsilon_{s}+\alpha(\varepsilon_{0}^{\prime}-\varepsilon_{0})}{kT}\right),\label{eq:73}
\end{equation}
where the pre-factor $\Lambda_{\varepsilon}$ absorbs the additional
entropy effects.

The chemical potential difference $(\mu_{0}^{\prime}-\mu_{0})$ is
proportional to the GB excess free energy (per unit area), $\gamma=u-T\eta$,
where $u$ is the GB energy and $\eta$ is the GB entropy. Representing
the GB by a uniform slab of width $\delta$, we have 
\begin{equation}
\gamma=\dfrac{(\mu_{0}^{\prime}-\mu_{0})\delta}{\Omega^{\prime}},\qquad u=\dfrac{(\varepsilon_{0}^{\prime}-\varepsilon_{0})\delta}{\Omega^{\prime}}.\label{eq:74}
\end{equation}
(Recall that $\Omega^{\prime}$ is the atomic volume in the GB.) Following
Borisov et al.~\citep{Borisov}, these equations can be recast in
the form of 
\begin{equation}
\gamma=\dfrac{(\mu_{0}^{\prime}-\mu_{0})m}{a^{2}},\qquad u=\dfrac{(\varepsilon_{0}^{\prime}-\varepsilon_{0})m}{a^{2}},\label{eq:76}
\end{equation}
where $m$ is the number of parallel atomic layers in the GB and $a=\left(\Omega^{\prime}\right)^{1/3}$
is the approximate distance between atoms. Equations (\ref{eq:72})
and (\ref{eq:73}) become, respectively, 
\begin{equation}
\dfrac{D^{\prime}}{D}=\Lambda_{g}\exp\left(\dfrac{g_{s}}{kT}\right)\exp\left(\dfrac{\alpha a^{2}\gamma}{mkT}\right),\label{eq:77}
\end{equation}
\begin{equation}
\dfrac{D^{\prime}}{D}=\Lambda_{\varepsilon}\exp\left(\dfrac{\varepsilon_{s}}{kT}\right)\exp\left(\dfrac{\alpha a^{2}u}{mkT}\right).\label{eq:78}
\end{equation}
Qualitatively, these equations predict that diffusion in high-energy
GBs must be faster than diffusion in low-energy GBs and that strong
GB segregation of an alloy component can reduce the GB diffusion coefficient
of that component (trapping effect). 

The importance of Eq.(\ref{eq:72}) is that it unifies, in a single
and simple equation, the three most fundamental properties of GBs:
GB free energy, GB segregation, and GB diffusion. This equation establishes
a link between the thermodynamic and kinetic properties of GBs. Experimental
methods are available for reasonably accurate measurements of GB diffusion
coefficients and segregation factors \citep{Kaur95,Balluffi95}. However,
experimental measurements of GB free energies are indirect, unreliable,
and rare. Atomistic calculations of $\gamma$ are also challenging.
Equation (\ref{eq:72}) enables predictions of GB free energies from
experimental data on GB diffusion and GB segregation.

Predictions from the above equations must be taken with great caution.
This equation is based on many approximations whose significance is
not fully understood and requires further analysis. For example, the
equations describe GBs using fixed geometric parameters (such as $a$
and $\delta$) that in reality depend on the GB character. Furthermore,
describing the GB segregation by a single segregation free energy
$g_{s}$ does not capture the well-established spectrality of GB segregation
\citep{Matson:2026aa}. These and other model approximations can affect
its comparison with experimental data. Nevertheless, these equations
can provide useful complementary information that can be compared
with available experimental data and direct computer simulations.

Equation (\ref{eq:77}) can be recast in the form 
\begin{equation}
\dfrac{s\delta D^{\prime}}{D}=\delta\Lambda_{g}\exp\left(\dfrac{\alpha a^{2}\gamma}{mkT}\right),\label{eq:77-1}
\end{equation}
or in the energy representation,

\begin{equation}
\dfrac{s\delta D^{\prime}}{D}=\delta\Lambda_{\varepsilon}\exp\left(\dfrac{\alpha a^{2}u}{mkT}\right).\label{eq:78-1}
\end{equation}
Here, 
\begin{equation}
s\equiv\dfrac{c_{s}^{\prime}}{c_{s}}=\exp\left(-\dfrac{g_{s}}{kT}\right)\label{eq:78-2}
\end{equation}
is the GB segregation factor defined as the ratio of the solute concentrations
in the GB ($c_{s}^{\prime}$) and in the lattice ($c_{s}$). Equations
(\ref{eq:77-1}) and (\ref{eq:78-1}) establish a link with GB diffusion
experiments, most of which measure the triple product $P=s\delta D^{\prime}$
\citep{Kaur95}. Note that equations (\ref{eq:77-1}) and (\ref{eq:78-1})
were obtained assuming a linear segregation isotherm in Eq.(\ref{eq:78-2}),
which is only valid in the limit of infinite dilution. In experimental
measurements, the triple product $P$ is likewise extracted from the
diffusion penetration profiles under the assumption that $s$ remains
constant along the GB, which is equivalent to a linear segregation
isotherm.

When the above equations are applied to diffusion and segregation
of substitutional impurity atoms, $\gamma$ and $u$ represent the
GB free energy and energy in the pure solvent. In other words, substitutional
impurity diffusion can be used as a probe to estimate the GB free
energy in the host material. For impurity atoms diffusing by the direct
interstitial mechanism, the situation is more complex. Under the reasonable
assumption of $n=1$, the diffusivity ratio $D^{\prime}/D$ does not
depend on GB free energy. Such impurities are unsuitable for extracting
the GB free energy from diffusion data.

\subsection{Comparison with literature}

\subsubsection{Comparison with Borisov et al.~\citep{Borisov}}

An equation similar to Eq.(\ref{eq:78}), but without the segregation
factor, was derived by Borisov et al.~\citep{Borisov} for self-diffusion
in elemental systems. For comparison with their equations, consider
the particular case of Eq.(\ref{eq:78}) when $\varepsilon_{s}=0$
and $\alpha=2$ (e.g., vacancy self-diffusion with $n=1$). Using
the relevant equations from sections \ref{subsec:TST} and \ref{subsec:Diffusion-kinetics},
we can express the pre-factor $\Lambda_{\varepsilon}$ in Eq.(\ref{eq:78})
in the following form: 
\begin{equation}
\Lambda_{\varepsilon}=\dfrac{f^{\prime}\xi^{\prime}l^{\prime2}}{f\xi l^{2}}\left(\dfrac{{\displaystyle {\displaystyle \prod_{i=1}^{3N^{\prime}}\nu_{i}^{\prime}}}}{{\displaystyle \prod_{i=1}^{3N}\nu_{i}}}\right)^{2}\left\{ {\displaystyle \dfrac{{\displaystyle \prod_{i=1}^{3N-1}\nu_{i}^{*}}}{{\displaystyle \prod_{i=1}^{3N^{\prime}-1}\nu_{i}^{*\prime}}}}\exp\left(-\dfrac{\hat{\varepsilon}^{\prime}-\hat{\varepsilon}}{kT}\right)\right\} .\label{eq:84}
\end{equation}
Borisov et al.~\citep{Borisov} derived their equation starting from
the relation $D^{\prime}/D=(\tau/\tau^{\prime})^{\alpha}$, where
$\tau$ and $\tau^{\prime}$ are the residence times of atoms in the
lattice and in the GB, respectively. Their equation for $D^{\prime}/D$
is (in our notation)\footnote{The original paper by Borisov et al.~\citep{Borisov} had a factor
of 2 inconsistency in the definition of the GB width $\delta$ in
their equation (5), which was pointed out and corrected by Pelleg
\citep{Pelleg:1966aa}, and is corrected in our analysis.} 
\begin{equation}
\dfrac{D^{\prime}}{D}=\lambda^{\alpha}\exp\left(\dfrac{\alpha a^{2}u}{mkT}\right),\label{eq:81}
\end{equation}
where they defined $\lambda=\tau/\tau^{\prime}$. For the vacancy
mechanism, they take $\alpha=2$ and their version of the pre-factor
$\Lambda_{\varepsilon}$ is 
\begin{equation}
\lambda^{2}=\left(\dfrac{{\displaystyle {\displaystyle \prod_{i=1}^{3N^{\prime}}\nu_{i}}}}{{\displaystyle \prod_{i=1}^{3N}\nu_{i}}}\right)^{2}\left\{ {\displaystyle \dfrac{{\displaystyle \prod_{i=1}^{3N-1}\nu_{i}^{*}}}{{\displaystyle \prod_{i=1}^{3N^{\prime}-1}\nu_{i}^{*\prime}}}}\exp\left(-\dfrac{\hat{\varepsilon}^{\prime}-\hat{\varepsilon}}{kT}\right)\right\} ^{2}.\label{eq:82}
\end{equation}
Comparing this equation with our Eq.(\ref{eq:84}), we see that they
omitted the jump correlation factors and structural differences between
the GB and the lattice, which are represented by the variables $l$,
$l^{\prime}$, $\xi$, and $\xi^{\prime}$. In addition, the expression
in the curly braces is squared in their equation and not squared in
ours. As discussed in the Appendix, the relation $D^{\prime}/D=(\tau/\tau^{\prime})^{2}$
used by Borisov et al.~\citep{Borisov} is invalid, which explains
the above discrepancy.

The key assumption of the Borisov model is that the activated complexes
of diffusive events in the GB and in the lattice are the same or at
least very close to each other. Borisov et al.~\citep{Borisov} formulated
this assumption as the equality of the energies $\hat{\varepsilon}^{\prime}$
and $\hat{\varepsilon}$. Under this condition, the exponential factor
with $(\hat{\varepsilon}^{\prime}-\hat{\varepsilon})$ becomes unity.
However, the difference between the two equations still remains due
to the pre-factors with $\nu_{i}^{*\prime}$ and $\nu_{i}^{*}$, which
arise from the vibrational free energies of the transition states
in the GB and in the lattice. However, note that the expression in
curly braces is equal to $\exp(-(\hat{g}^{\prime}-\hat{g})/kT)$.
Therefore, if we reformulate the said assumption as the equality of
the \emph{free} energies of the activated complexes, not simply energies,
then $\hat{g}^{\prime}=\hat{g}$ and the factor in the curly braces
becomes unity, making equations (\ref{eq:82}) and (\ref{eq:84})
identical (again, disregarding the first factor related to the correlation
factors and structural effects). We hypothesize that Borisov et al.~\citep{Borisov}
did not consider the vibrational pre-factor important and dropped
it. This assumption is equivalent to neglecting the difference between
the energies and free energies of the activated complexes.

As noted above, Borisov et al.~\citep{Borisov} considered only self-diffusion
of host atoms, not diffusion of solute atoms. Their theory was demonstrated
by application to Fe self-diffusion by the vacancy mechanism in $\gamma$
(FCC) and $\alpha$ (BCC) iron, as well as Fe-B alloys of both phases
assuming $\lambda=1$ and $m=1$. According to their estimates, boron
reduces the GB energy in both phases.

The most significant discrepancy between the equations derived by
Borisov et al.~\citep{Borisov} and the present analysis is related
to diffusion by the interstitial mechanism, for which they applied
Eq.(\ref{eq:81}) with $\alpha=1$. They did not specify whether they
considered direct or indirect diffusion mechanism. However, their
equation $D^{\prime}/D=\tau/\tau^{\prime}$ was justified by not requiring
the presence of point defects for the atomic jumps, which points to
the direct hopping mechanism. The direct hopping mechanism does not
operate for self-interstitials in the perfect lattice. Even if it
did, the value of $\alpha$ in Eq.(\ref{eq:81}) would be $\alpha=0$,
not $\alpha=1$, as was shown in section \ref{subsec:direct-interst}.

As discussed in the Appendix, the equation $D^{\prime}/D=\tau/\tau^{\prime}$
itself is meaningful, but the calculation of $\tau/\tau^{\prime}$
for the direct interstitial mechanism published by Borisov et al.~\citep{Borisov}
was based on the incorrect assumption that the free energy of interstitial
atoms in the GB was shifted relative to their free energy in the lattice
by the amount of $(\mu_{0}^{\prime}-\mu_{0})>0$. In fact, the shift
is $(\tilde{\mu}_{I}^{\prime}-\tilde{\mu}_{I})$ and is most likely
negative because the interstitial concentration in GBs is usually
higher than in the lattice. This leads to the following conclusions:
(1) Assuming that the free energy of the activated complex is nearly
the same in the GB and in the lattice ($\hat{g}^{\prime}\approx\hat{g}$),
the interstitial atoms in the GB are further away from the activated
state than the interstitial atoms in the lattice, leading to slower
diffusion of interstitial atoms in the GB relative to the lattice.
(2) The diffusivity ratio $D^{\prime}/D$ cannot be expressed as a
function of $(\mu_{0}^{\prime}-\mu_{0})$ and therefore the GB free
energy. The latter cannot be extracted from the diffusion data as
in the case of vacancy-mediated diffusion. These conclusions apply
equally to self-diffusion and diffusion of impurity atoms by the direct
interstitial mechanism.

\subsubsection{Various forms of the Borisov equation}

\label{subsec:Various-forms}

In most applications, the model proposed by Borisov et al.~\citep{Borisov}
has been used to extract the GB free energy $\gamma$ or GB energy
$u$ (a distinction is rarely made) using experimental data for GB
diffusion. Equation (\ref{eq:77-1}) is transformed into an application-ready
form by solving for $\gamma$: 
\begin{equation}
\gamma=\dfrac{mkT}{\alpha a^{2}}\ln\left(\dfrac{sD^{\prime}}{\Lambda_{g}D}\right).\label{eq:99}
\end{equation}

Assuming that the diffusion coefficients follow the Arrhenius relations
\begin{equation}
D=D_{0}\exp\left(-\dfrac{Q}{kT}\right),\;\;D^{\prime}=D_{0}^{\prime}\exp\left(-\dfrac{Q^{\prime}}{kT}\right),\label{eq:98}
\end{equation}
Eq.(\ref{eq:99}) becomes 
\begin{equation}
\gamma=\dfrac{m}{\alpha}\dfrac{Q-Q^{\prime}}{a^{2}}+\dfrac{mkT}{\alpha a^{2}}\ln\left(\dfrac{sD_{0}^{\prime}}{\Lambda_{g}D_{0}}\right).\label{eq:97}
\end{equation}
Here $Q$ and $Q^{\prime}$ are the activation enthalpies of the lattice
and GB diffusion, and $D_{0}$ and $D_{0}^{\prime}$ are the respective
pre-exponential factors. Assuming also that GB segregation follows
the Arrhenius relation 
\begin{equation}
s=s_{0}\exp\left(-\dfrac{Q_{s}}{kT}\right),\label{eq:96}
\end{equation}
where $Q_{s}<0$ is the segregation enthalpy and $s_{0}$ the segregation
pre-factor, Eq.(\ref{eq:97}) takes the form 
\begin{equation}
\gamma=\dfrac{m}{\alpha}\dfrac{Q-Q^{\prime}-Q_{s}}{a^{2}}+\dfrac{mkT}{\alpha a^{2}}\ln\left(\dfrac{s_{0}D_{0}^{\prime}}{\Lambda_{g}D_{0}}\right).\label{eq:95}
\end{equation}

It was proposed \citep{GUIRALDENQ,Gupta:1977aa} to write the last
equation as $\gamma=H-TS$ and interpret 
\begin{equation}
H=\dfrac{m}{\alpha}\dfrac{Q-Q^{\prime}-Q_{s}}{a^{2}}\label{eq:94}
\end{equation}
as the GB enthalpy and 
\begin{equation}
S=\dfrac{mk}{\alpha a^{2}}\ln\left(\dfrac{\Lambda_{g}D_{0}}{s_{0}D_{0}^{\prime}}\right)\label{eq:93}
\end{equation}
as the GB entropy (both per unit area). Equations (\ref{eq:99})-(\ref{eq:93})
appear in the literature in many different notation and usually with
$\Lambda_{g}=\lambda^{\alpha}$ \citep{Pelleg:1966aa,GUIRALDENQ,Gupta:1977aa}.
Due to the challenge in quantifying the parameters appearing in the
above equations, most publications assume $m=1$ and $\lambda=1$.

\subsubsection{Comparison with self-diffusion experiments}

\label{subsec:Comparison-with-self-diffusion}

The GB free energies extracted from experimental diffusion data often
have a reasonable order of magnitude. Gupta published a comparison
between the measured and predicted GB free energies for five FCC metals,
showing a significant correlation \citep{Gupta:1977aa}. Predictions
often reproduce general trends, such as a higher energy of high-angle
GBs compared to low-angle and special GBs \citep{Li:2023ab}. The
latter trend already follows from Eq.(\ref{eq:94}) because in most
materials $Q>Q^{\prime}$ except for low-angle and some special GBs,
for which $Q\apprge Q^{\prime}$ \citep{Kaur95}. According to Eq.(\ref{eq:97}),
the same trend should be followed by the GB enthalpy $H$. We caution,
however, against over-reliance on direct comparisons with experimental
data when testing the model capabilities. The experimental GB free
energies come primarily from grooving experiments whose reliability
is limited due to challenges in ascertaining the groove equilibrium,
errors in measuring the groove angles, and the scarcity of independently
measured surface energies required for calculating the GB energies.

Experiments and atomistic simulations \citep{Zhu:2018aa,MRS-Bulletin-GB-phases,Frolov2012b}
indicate that GB free energy decreases with temperature primarily
due to structural disordering at high temperatures. This well-known
trend was reproduced by several authors using the Borisov equation
with input from self-diffusion measurements \citep{Borisov,Pelleg:1966aa,GUIRALDENQ,Chen:2003aa,Divinski:2010aa,Li:2023ab}.
This agreement cannot be interpreted as the ability of the model to
predict thermal disordering of the GB structure. The disorder effect
is not part of the model. According to this model, the decrease in
$\gamma$ with temperature can only come from a positive GB entropy
$S$ given by Eq.(\ref{eq:93}). This entropy is associated solely
with atomic vibrations. With the usual assumption of $\Lambda_{g}=1$
and taking $s=1$ (self-diffusion), the sign of $S$ depends on the
ratio of the pre-exponential factors $D_{0}/D_{0}^{\prime}$. There
is no general rule by which this ratio would be greater or smaller
than unity. It can vary from one material to another and depend on
the GB type and diffusion mechanism. It is empirically known that
for most metals, $D_{0}$ and $D_{0}^{\prime}$ are on the order of
$10^{-5}$ to $10^{-3}$ m$^{2}$/s and are not very different from
each other \citep{Kaur95,Mehrer2007}. There is also a significant
uncertainty in the experimental values of $D_{0}$ and $D_{0}^{\prime}$
as the Arrhenius law is not followed exactly, and calculations of
the pre-exponential factors require extrapolation far outside the
temperature interval of measurements. In addition, as already mentioned,
most experimental measurements yield only the product $\delta D^{\prime}$.
The value of $D_{0}^{\prime}$ obtained from the measurements depends
on the assumption about the GB width $\delta$. The widely accepted
value is $\delta=0.5$ nm \citep{Kaur95}, but experimental and simulation
estimates vary approximately between 0.5 and 1.5 nm; see, for example,
\citep{Prokoshkina:2013aa,Tuchinda:2024aa}.

Furthermore, the basic assumption of the model is that the diffusion
mechanism, the structure, and the jump correlation factor are identical
in the GB and in the lattice. Under this assumption, there is no basis
to expect that one pre-exponential factor would be systematically
larger than the other. Taking into account these uncertainties, it
is natural to expect that the value of $S$ predicted by Eq.(\ref{eq:93})
can be positive, negative, or nearly zero. In fact, in several metals,
the GB free energy extracted from the Borisov equation increased with
temperature or did not change much \citep{Pelleg:1966aa,GUIRALDENQ}.
Of course, one can always argue that such cases were affected by uncontrolled
impurities that altered the GB free energy \citep{Divinski:2010aa,Prokoshkina:2013aa}.
Overall, it is not obvious why the model would systematically predict
a positive GB entropy. This aspect of the Borisov model is not well-understood
and warrants a more detailed investigation.

GB free energies were also extracted from self-diffusion measurements
on materials containing segregating solutes. The GB free energies
were typically lower than in the absence of segregation and often
increased with temperature. This effect was reported already in the
original publication by Borisov et al.~\citep{Borisov} for Fe self-diffusion
in Fe-B alloys. Their work was followed by many other studies published
by Pelleg \citep{Pelleg:1966aa}, Guiraldenq \citep{GUIRALDENQ},
Gupta \citep{Gupta:1977aa}, and others \citep{Chen:2003aa,Lin:2017aa}.
More recently, Divinski et al.~\citep{Divinski:2010aa,Prokoshkina:2013aa}
measured GB self-diffusion in polycrystalline Ni of varying purity
and extracted the effective GB free energy as a function of temperature.
In high-purity Ni, the GB free energy was larger and decreased with
temperature, whereas in less pure Ni, it was lower and slightly increased
with temperature.

The results mentioned above are consistent with interface thermodynamics.
According to the Gibbs adsorption equation \citep{Gibbs}, solute
atoms segregate to interfaces to reduce their free energy. The amount
of segregation and therefore the reduction in free energy are highest
at low temperatures and decrease with increasing temperature, leading
to a positive slope of the plot of $\gamma$ versus temperature. This
trend competes with the opposing trend of decreasing $\gamma$ with
temperature due to structural disorder. If the segregation effect
is strong enough, the first trend wins, resulting in a positive slope.

Note that the solute segregation parameters do not appear in the self-diffusion
version of the Borisov equation explicitly. However, they are captured
by the self-diffusion measurements and affect the GB free energies
predicted by that equation. As a result, comparison of the GB free
energies obtained from self-diffusion measurements on chemically pure
and segregated samples can yield information about the solute segregation.
Specifically, the GB segregation parameters can be calculated by measuring
the difference between the GB free energies with GB segregation ($\gamma$)
and without segregation ($\gamma_{0}$) and using the Gibbs adsorption
equation.

For example, Gupta \citep{Gupta:1977aa} applied the McLean-Langmuir
segregation isotherm \citep{McLean}, 
\begin{equation}
\dfrac{c_{s}^{\prime}}{1-c_{s}^{\prime}}\dfrac{1-c_{s}}{c_{s}}=\exp\left(-\dfrac{g_{s}}{kT}\right),\label{eq:142}
\end{equation}
where $c_{s}^{\prime}$ and $c_{s}$ are the fractions of available
sites occupied by the solute atoms in the GB and in the lattice, respectively.
Inserting this isotherm into the Gibbs adsorption equation and integrating
from the pure solvent to the binary solution, it is easy to obtain
\begin{equation}
\gamma(T)-\gamma_{0}(T)=-\xi kT\ln\left[1-c_{s}+c_{s}\exp\left(-\dfrac{g_{s}(T)}{kT}\right)\right],\label{eq:143}
\end{equation}
where $\xi$ denotes the number of available adsorption sites per
unit GB area. Knowing the functions $\gamma(T)$ and $\gamma_{0}(T)$,
one can solve this equation for the segregation free energy $g_{s}$
as a function of temperature. Alternatively, Eq.(\ref{eq:143}) can
be written in the form 
\begin{equation}
c_{s}\exp\left(-\dfrac{g_{s}(T)}{kT}\right)=c_{s}-1+\left[\dfrac{D^{\prime}}{D}\left(\dfrac{D}{D^{\prime}}\right)_{p}\right]^{1/2},\label{eq:143-1}
\end{equation}
where the index $p$ refers to the pure solvent. We used the relation
$\xi=m/a^{2}$ and assumed self-diffusion by the vacancy mechanisms
($\alpha=2$). Equations (\ref{eq:143}) and (\ref{eq:143-1}) allow
calculations of the segregation energy and segregation entropy from
diffusion data.

Gupta was successful in estimating segregation parameters in a number
of alloy systems from radiotracer self-diffusion measurements. This
method appears to be quite powerful. However, it should be noted that
it attributes the temperature dependence of $\gamma$ entirely to
the GB segregation effect, disregarding the possible effect of GB
disordering at high temperatures, which can also contribute to the
temperature dependence of $\gamma$. 

It should also be noted that the McLean-Langmuir segregation isotherm
in Eq.(\ref{eq:142}) is valid only under the assumptions of perfect
solutions in the GB and in the grains and a single segregation free
energy. The reader is cautioned against applying this equation to
concentrated alloys, where solute-solute interactions in the GBs can
be significant, causing strong deviations from Eq.(\ref{eq:142})
and even phase separation within the GBs. Gupta's experiments were
on dilute alloys containing only one or less than one atomic percent
of the solute. Even then, the spectrality of GB segregation energies
\citep{Matson:2026aa} may call into question the concept of a single
segregation free energy in Eq.(\ref{eq:142}).

\subsubsection{Applications to impurity diffusion}

In section \ref{subsec:Impurity-diffusion}, we derived a generalized
version of the Borisov model \citep{Borisov} applicable to impurity
diffusion; see equations (\ref{eq:77}), (\ref{eq:99}) and the subsequent
equations in section \ref{subsec:Various-forms}. For substitutional
impurity atoms, this model can be applied with $n=1$ ($\alpha=2$)
to calculate the GB free energy from impurity diffusion data. As indicated
above, experimental measurements of substitutional impurity diffusion
can be used as a probe of GB free energy in the host material assuming
that the segregation factor has been measured independently. To our
knowledge, this opportunity has not been realized so far.

Several authors measured GB diffusion of substitutional elements and
used the results to calculate GB free energy. However, in most cases,
these elements were used essentially as proxies for self-diffusion
of the solvent. For example, Chen et al.~\citep{Chen:2003aa} measured
lattice and GB diffusion of Cr in an Inconel alloy with the chemical
composition Ni-16\%Cr-7\%Fe (mass percents) using a radioactive isotope
of Cr. Assuming the vacancy mechanism, they extracted the GB free
energies from the Borisov equation. GB segregation of Cr was not considered.
Instead, the discussion focused on the role of carbon impurity atoms
whose concentration was at a level of 0.07\%. Carbon was shown to
decrease the GB free energy, but its diffusion was not measured. Lin
et al.~\citep{Lin:2017aa} back-calculated Cr GB diffusion coefficients
from the oxidation kinetics of a Fe-22wt.\%Cr alloy. The Borisov equation
was used assuming the vacancy diffusion mechanism and $s=1$ (no Cr
segregation). The obtained GB energies were substantially lower than
those predicted by direct atomistic simulations but followed a similar
trend as a function of disorientation. Here again, the segregation
factor was not considered, and Cr diffusion was treated as if it were
self-diffusion of the solvent (Fe). A similar approach was used in
recent studies of GB diffusion in high-entropy alloys \citep{Vaidya:2017aa,Glienke:2020aa,Choi:2024aa,Yadav:2026aa}.

It is important to emphasize the difference between the two scenarios
involving GB segregation of solute atoms: 
\begin{itemize}
\item In one scenario, discussed in section \ref{subsec:Comparison-with-self-diffusion},
self-diffusion of the solvent element is measured in the presence
of GB segregation of a solute element (or elements). Information about
solute segregation is not used during the calculation of the GB free
energy. The Borisov equation is applied with $s=1$. However, because
the solute segregation affects the GB diffusion coefficient, the segregation
parameters can be evaluated by comparing the obtained GB free energies
with those extracted from GB diffusion in the pure solvent. The calculation
of the segregation parameters is then based on equations (\ref{eq:143})
and (\ref{eq:143-1}). To use this approach, the segregation isotherm
does not have to be linear. 
\item In the second scenario, GB diffusion of solute atoms is measured,
and the GB free energy is obtained from Eq.(\ref{eq:99}) or one of
its versions appearing in section \ref{subsec:Various-forms}. This
calculation requires knowledge of the segregation factor $s$. It
is assumed that the solute is substitutional and that GB segregation
is linear. 
\end{itemize}
It would be interesting to do both: measure GB self-diffusion of the
solvent and GB impurity diffusion in the same material. The segregation
factor calculated from the self-diffusion measurements by Gupta's
method can be fed into the calculation of GB free energy based on
the impurity diffusion measurements. This would enable a cross-check
of the Borisov model by comparing the GB free energies obtained by
the two methods. This scheme requires the impurity to be substitutional
and dilute because the impurity diffusion measurements are based on
the assumption of a constant GB segregation factor \citep{Kaur95}.

For impurity atoms diffusing by the direct interstitial mechanism,
the situation is more complex. Under the reasonable assumption of
$n=1$, the impurity diffusivity ratio $D^{\prime}/D$ is independent
of the GB free energy, which precludes the extraction of the GB free
energy from the diffusion data. Note that Eq.(\ref{eq:59-1}) predicts
a trapping effect, in which segregating impurities ($g_{s}<0$) diffuse
in GBs slower than in the lattice. This prediction relies on the assumption
of equal free energies of the activated complexes in the GB and in
the lattice. If the two free energies are not exactly equal, the trapping
effect will compete with the usual short-circuit effect that causes
an acceleration of GB diffusion. Depending on the outcome of the competition,
GB diffusion can be faster or slower than lattice diffusion. This
explains the variability of the diffusion data on hydrogen GB diffusion
in metals, which can be faster or slower than lattice diffusion \citep{Birnmaun-1988,Yao:1991aa,Harris:1991aa,Brass:1996aa,Pedersen2009,Oudriss:2012aa,Zhou:2019aa,Page:2021aa,Wang:2022ac}.

We emphasize that these considerations are based on the simplifying
assumption that GB diffusion occurs by the same mechanism as lattice
diffusion and is mediated by either vacancies or interstitials. Atomistic
modeling suggests that GB diffusion is more complex and is dominated
by collective mechanisms not found in the perfect lattice \citep{Mishin:2015ac,Suzuki05a,Chesser:2024aa}.
GB diffusion of interstitial elements, such as hydrogen, is likely
to occur by interstitial hops similar to those in the lattice ($\alpha=1$).
However, assigning a definitive $\alpha$ value to GB self-diffusion
and substitutional-solute diffusion can be problematic, especially
at high temperatures.

\subsection{Is the activated complex really ``universal''?}

Of many assumptions underlying the Borisov model \citep{Borisov},
the most critical one is the equality of the free energies of the
activated complexes in the GB and in the lattice. Testing the validity
of this assumption is important to understand the theoretical foundations
of the model.

The assumptions underlying the Borisov model were tested \emph{indirectly}
by assessing its predictive capacity. This was usually done by comparing
the model predictions for GB free energies with the experimental data.
As discussed above, the model reproduces the experimental data on
a semi-quantitative level and often captures the correct trends, such
as the temperature dependence of the GB free energy and its suppression
by impurity segregation. Although this agreement is reassuring, this
type of comparison hardly validates the specific activated complex
assumption, because the model relies on many other assumptions and
approximations.

A more \emph{direct} test can be designed by using atomistic computer
simulations with interatomic potentials. In potential-based simulations,
each atom is assigned an energy, allowing a direct calculation of
the energy of the atoms comprising an activated complex. It should
be straightforward to calculate the activation energies of many vacancy
jumps within the GB core as well as in the perfect lattice. This can
be accomplished by using the nudged elastic band method \citep{Jonsson98,HenkelmanJ00}
or similar techniques. The energy of the atom exchanging positions
with a vacancy when it reaches the saddle point along the minimum-energy
path is the energy of the activated complex.\footnote{We emphasize again that the energy of the activated complex must be
taken relative to isolated atoms, not relative to the energy of the
system with the vacancy before it jumps. The latter is the jump activation
barrier ($g^{*}$ or $g^{*\prime}$) and is a different property.} The latter can be computed for vacancy jumps in the GB ($\hat{g}^{\prime}$)
and in the lattice ($\hat{g}$). Similar calculations could be performed
for a substitutional or interstitial impurity atom. It is highly unlikely
that the obtained values of $\hat{g}$ and $\hat{g}^{\prime}$ will
be equal to each other. However, the calculations could estimate the
dispersion of these values compared to the variations in other energies
involved in the problem, such as the energies of the atoms before
the diffusive jumps and the point-defect formation energies. A relatively
narrow distribution of the energies of the activated complexes, if
confirmed, would justify the basic assumption of the model.

To our knowledge, the proposed computational test has not been conducted
so far. Urazaliev et al.~\citep{Urazaliev:2024aa} calculated self-diffusion
coefficients in a series of individual symmetrical tilt GBs in Ni
by MD simulations. They were able to distinguish between the vacancy
and interstitial diffusion mechanisms and extract the respective activation
energies for all GBs tested. The authors did not specify whether the
interstitial diffusion mechanism was direct or indirect, which makes
a significant difference as discussed above. They plotted the activation
energies against the directly computed GB energies and observed a
substantial scatter of the points without an obvious correlation.
It should be noted that their plot combined two different diffusion
mechanisms, making it difficult to draw conclusions. More importantly,
the activation energies shown in the plot are different from those
of the activated complexes. The two energies differ by the energy
of the respective defect-free structure, which is either the GB or
the lattice. Due to these uncertainties, no definitive conclusions
can be deduced from this study regarding the validity of the activated
complex assumption in the Borisov model.

Page et al.~\citep{Page:2021aa} investigated hydrogen diffusion
in 26 symmetrical tilt GBs in Ni by MD simulations with an embedded-atom
interatomic potential. The authors computed the GB and lattice diffusion
coefficients and applied Eq.(\ref{eq:99}) with $\alpha=1$ and $s=1$
to extract the GB energies. Hydrogen GB segregation was observed but
was not quantified or used in the calculations. The correlation with
the directly calculated GB energies was weak but could be improved
by treating $\lambda$ and $m$ as fitting parameters. As discussed
above, applications of the Borisov model to interstitial impurity
diffusion should be taken with a grain of salt. The model considers
only one type of interstitial sites, whereas hydrogen is known to
occupy both tetrahedral and octahedral sites in Ni (and other FCC
metals) with different solubility energies and jump barriers. More
importantly, since the appropriate value of $\alpha$ for interstitial
impurities is zero, the diffusivity ratio $D^{\prime}/D$ does not
depend on the GB free energy. Instead, the model predicts $D^{\prime}/D\propto1/s$
(see Eq.(\ref{eq:59-1})), i.e., slow GB diffusion due to the solute
trapping effect ($D^{\prime}<D$). The enhanced GB diffusion observed
in \citep{Page:2021aa} (and many other experimental and modeling
studies) signals a deviation from the Borisov model. However, we note
that the simulations in \citep{Page:2021aa} could have been useful
for testing the activated complex hypothesis if the hydrogen energies
in the activated states of diffusive jumps were calculated in the
GBs and in the lattice (relative to the same reference state).

\section{Summary and conclusions}

\label{sec:Summary}

The detailed derivation of the equation proposed by Borisov et al.~\citep{Borisov}
has allowed us to identify the assumptions underlying this model.
These assumptions can be summarized as follows: 
\begin{itemize}
\item The model considers a single GB represented by a uniform slab of width
$\delta$ composed of a fixed number $m$ of atomic planes. The number
of sites within the GB is fixed. 
\item Atoms diffuse in the GB and in the lattice by the \emph{same} mechanism. 
\item For the direct interstitial mechanism, the model assumes that there
is a single type of interstitial site in the GB and a single (possibly
different) type of interstitial site in the lattice. For example,
the model does not distinguish between the tetrahedral and octahedral
interstitial sites in the FCC and BCC crystal structures. For the
indirect interstitial mechanisms, the model neglects the difference
between collinear and non-collinear atomic rearrangements. 
\item The free energy of any collection of atoms includes their static potential
energy and their vibrational free energy in the harmonic approximation.
In the presence of point defects (e.g., vacancies, self-interstitials,
or impurity atoms), the model additionally includes the ideal configurational
entropy term $kT\ln c_{d}$, where $c_{d}\ll1$ is the point-defect
concentration. 
\item Structural differences between the GB and the lattice are ignored. 
\item Diffusive events are treated within the harmonic transition state
theory. 
\item The correlation factors between the atomic jumps in the GB and in
the lattice are assumed to be equal. 
\item In applications to a temperature interval, all model parameters are
assumed to be temperature-independent. 
\item Most crucially, it is assumed that all atomic rearrangements go through
an activated complex containing the same number $n$ of atoms and
having the same free energy. 
\end{itemize}
A clear distinction should be made between the cases of self-diffusion
and impurity diffusion. The original Borisov equation was derived
for self-diffusion in an elemental material or self-diffusion of solvent
atoms in an alloy system. We have shown that the model can be extended
to diffusion of a dilute solute component (impurity).

The generalized Borisov equation applicable to both self-diffusion
and impurity diffusion is 
\begin{equation}
\dfrac{sD^{\prime}}{D}=\Lambda_{g}\exp\left(\dfrac{\alpha a^{2}\gamma}{mkT}\right),\label{eq:77-1-1}
\end{equation}
where $\alpha=n+1$ for the vacancy diffusion mechanism and $\alpha=n-1$
for the interstitial diffusion mechanism. Assuming single atom jumps,
$\alpha=2$ for the vacancy mechanism and $\alpha=0$ for the direct
interstitial mechanism. In the latter case, the diffusivity ratio
does not depend on the GB free energy $\gamma$. For the indirect
interstitial mechanisms, the size of the activated complex depends
on the detailed atomic rearrangement during diffusive events, the
most typical values being $n=2$ (interstitialcy mechanism) and $n=3$
(dumbbell mechanism). For impurity diffusion, the equation includes
the GB segregation factor given by $s=\exp(-g_{s}/kT)$, where the
segregation free energy $g_{s}$ is negative. For self-diffusion,
the equation is used with $s=1$. These case are summarized in Table
\ref{tab:Summary}.

\begin{table}
\caption{Summary of the conditions under which the Borisov model can be applied.
}\label{tab:Summary}
\begin{tabular}{|l|l|c|c|c|}
\hline 
Diffusion type & Mechanism & $\alpha$ & Reasonable $n$ & $\gamma$ can be obtained?\tabularnewline
\hline 
\hline 
Self-diffusion & Vacancy & $n+1$ & 1 & Yes\tabularnewline
\hline 
Self-diffusion & Indirect interstitial & $n-1$ & 2 or 3 & Yes\tabularnewline
\hline 
Impurity diffusion & Vacancy & $n+1$ & 1 & Yes\tabularnewline
\hline 
Impurity diffusion & Direct interstitial & $n-1$ & 1 & No\tabularnewline
\hline 
\end{tabular}

\end{table}

The pre-exponential factor $\Lambda_{g}$ in Eq.(\ref{eq:77-1-1})
depends on vibrational frequencies and jump correlation factors. In
applications, $\Lambda_{g}$ is treated as a constant, which is usually
assumed to be $\Lambda_{g}=1$ for the lack of better estimates. For
the same reason, it is usually assumed that $m=1$. Perhaps a better
approach would be to treat $\Lambda_{g}$ and $m$ as calibration
parameters, as suggested by Page et al.~\citep{Page:2021aa}.

Equation (\ref{eq:77-1-1}) links the GB and lattice diffusion coefficients,
the GB segregation factor, and the GB free energy into a single and
simple functional form. Given the numerous assumptions and approximations
made during its derivation, this equation is not expected to be very
accurate. However, it can be considered a useful empirical correlation.
Remarkably, this correlation works fairly well in many cases, often
predicting reasonable values of GB energies and reproducing important
trends related to the temperature and impurity effects (see discussions
in the previous section). When using the Borisov equation, attention
should be paid to the correct choice of the $\alpha$ and $s$ values
according to the assumed diffusion mechanism and the type of diffusion
measurements (self-diffusion versus impurity diffusion).

On the theoretical side, it is important to test the validity of the
activated complex hypothesis, which is the main pillar of the Borisov
model. As discussed above, this can be accomplished by specially designed
molecular statics simulations. Ideally, separate tests should be conducted
for self-diffusion and impurity diffusion. There are no fundamental
obstacles to such tests. \\

\noindent\textbf{Acknowledgements}

\noindent This research was supported by the U.S.~Department of Energy,
Office of Basic Energy Sciences, Division of Materials Sciences and
Engineering, under Award \# DE-SC0023102.


\expandafter\ifx\csname natexlab\endcsname\relax
\global\long\def\natexlab#1{#1}%
\fi \providecommand{\url}[1]{\texttt{#1}} \providecommand{\href}[2]{#2}
\providecommand{\path}[1]{#1} \providecommand{\DOIprefix}{doi:}
\providecommand{\ArXivprefix}{arXiv:} \providecommand{\URLprefix}{URL: }
\providecommand{\Pubmedprefix}{pmid:} \providecommand{\doi}[1]{\href{http://dx.doi.org/#1}{\path{#1}}}
\providecommand{\Pubmed}[1]{\href{pmid:#1}{\path{#1}}} \providecommand{\bibinfo}[2]{#2}
\ifx\xfnm\relax \def\xfnm [#1]{\unskip,\space#1}\fi 
 

\begin{thebibliography}{47}
    \expandafter\ifx\csname natexlab\endcsname\relax\def\natexlab#1{#1}\fi
    \providecommand{\url}[1]{\texttt{#1}}
    \providecommand{\href}[2]{#2}
    \providecommand{\path}[1]{#1}
    \providecommand{\DOIprefix}{doi:}
    \providecommand{\ArXivprefix}{arXiv:}
    \providecommand{\URLprefix}{URL: }
    \providecommand{\Pubmedprefix}{pmid:}
    \providecommand{\doi}[1]{\href{http://dx.doi.org/#1}{\path{#1}}}
    \providecommand{\Pubmed}[1]{\href{pmid:#1}{\path{#1}}}
    \providecommand{\bibinfo}[2]{#2}
    \ifx\xfnm\relax \def\xfnm[#1]{\unskip,\space#1}\fi
    \bibitem[{Borisov et~al.(1964)Borisov, Golikov, and Scherbedinsky}]{Borisov}
    \bibinfo{author}{V.~T. Borisov}, \bibinfo{author}{V.~M. Golikov},
    \bibinfo{author}{G.~V. Scherbedinsky},
    \newblock \bibinfo{title}{Relation between diffusion coefficients and grain
        boundary energy},
    \newblock \bibinfo{journal}{Phys. Met. Metallogr.} \bibinfo{volume}{17}
    (\bibinfo{year}{1964}) \bibinfo{pages}{881--885}.
    \bibitem[{Larch{\'e} and Cahn(1978)}]{Larche_Cahn_78}
    \bibinfo{author}{F.~C. Larch{\'e}}, \bibinfo{author}{J.~W. Cahn},
    \newblock \bibinfo{title}{Thermodynamic equilibrium of multiphase solids under
        stress},
    \newblock \bibinfo{journal}{Acta Metall.} \bibinfo{volume}{26}
    (\bibinfo{year}{1978}) \bibinfo{pages}{1579--1589}.
    \bibitem[{Larch{\'e} and Cahn(1973)}]{Larche73}
    \bibinfo{author}{F.~Larch{\'e}}, \bibinfo{author}{J.~W. Cahn},
    \newblock \bibinfo{title}{A linear theory of thermochemical equilibrium of
        solids under stress.},
    \newblock \bibinfo{journal}{Acta Metall.} \bibinfo{volume}{21}
    (\bibinfo{year}{1973}) \bibinfo{pages}{1051--1063}.
    \bibitem[{Larch{\'e} and Cahn(1985)}]{Larche1985}
    \bibinfo{author}{F.~C. Larch{\'e}}, \bibinfo{author}{J.~W. Cahn},
    \newblock \bibinfo{title}{The interactions of composition and stress in
        crystalline solids},
    \newblock \bibinfo{journal}{Acta Metall.} \bibinfo{volume}{33}
    (\bibinfo{year}{1985}) \bibinfo{pages}{331--367}.
    \bibitem[{Mullins and Sekerka(1985)}]{Mullins1985}
    \bibinfo{author}{W.~W. Mullins}, \bibinfo{author}{R.~F. Sekerka},
    \newblock \bibinfo{title}{On the thermodynamics of crystalline solids},
    \newblock \bibinfo{journal}{J. Chem. Phys.} \bibinfo{volume}{82}
    (\bibinfo{year}{1985}) \bibinfo{pages}{5192--5202}.
    \bibitem[{Eyring(1935)}]{Eyring:1935aa}
    \bibinfo{author}{H.~Eyring},
    \newblock \bibinfo{title}{The activated complex in chemical reactions},
    \newblock \bibinfo{journal}{The Journal of Chemical Physics}
    \bibinfo{volume}{3} (\bibinfo{year}{1935}) \bibinfo{pages}{107--115}.
    \bibitem[{Vineyard(1957)}]{Vineyard:1957vo}
    \bibinfo{author}{G.~H. Vineyard},
    \newblock \bibinfo{title}{Frequency factors and isotope effects in solid state
        rate processes},
    \newblock \bibinfo{journal}{Journal of Physics and Chemistry of Solids}
    \bibinfo{volume}{3} (\bibinfo{year}{1957}) \bibinfo{pages}{121--127}.
    \bibitem[{Philibert(1991)}]{Philibert}
    \bibinfo{author}{J.~Philibert}, \bibinfo{title}{Atom Movements -- Diffusion and
        Mass Transport in Solids}, \bibinfo{publisher}{Les Editions de Physique},
    \bibinfo{address}{Les Ulis}, \bibinfo{year}{1991}.
    \bibitem[{Mehrer(2007)}]{Mehrer2007}
    \bibinfo{author}{H.~Mehrer}, \bibinfo{title}{Diffusion in Solids: Fundamentals,
        Methods, Materials}, \bibinfo{publisher}{Springer Verlag},
    \bibinfo{address}{Berlin}, \bibinfo{year}{2007}.
    \bibitem[{Borisov et~al.(1963)Borisov, Golikov, and
        Shcherbedinskiy}]{Borisov_1963}
    \bibinfo{author}{V.~T. Borisov}, \bibinfo{author}{V.~M. Golikov},
    \bibinfo{author}{G.~V. Shcherbedinskiy},
    \newblock \bibinfo{title}{Statistical calculation of self-diffusion coefficient
        in metals},
    \newblock \bibinfo{journal}{Doklady Akademii nauk SSSR} \bibinfo{volume}{149}
    (\bibinfo{year}{1963}) \bibinfo{pages}{1307--1310}.
    \bibitem[{Kaur et~al.(1995)Kaur, Mishin, and Gust}]{Kaur95}
    \bibinfo{author}{I.~Kaur}, \bibinfo{author}{Y.~Mishin},
    \bibinfo{author}{W.~Gust}, \bibinfo{title}{Fundamentals of Grain and
        Interphase Boundary Diffusion}, \bibinfo{publisher}{Wiley},
    \bibinfo{address}{Chichester, West Sussex}, \bibinfo{year}{1995}.
    \bibitem[{Sutton and Balluffi(1995)}]{Balluffi95}
    \bibinfo{author}{A.~P. Sutton}, \bibinfo{author}{R.~W. Balluffi},
    \bibinfo{title}{Interfaces in Crystalline Materials},
    \bibinfo{publisher}{Clarendon Press}, \bibinfo{address}{Oxford},
    \bibinfo{year}{1995}.
    \bibitem[{Matson et~al.(2026)Matson, Tuchinda, and Schuh}]{Matson:2026aa}
    \bibinfo{author}{T.~P. Matson}, \bibinfo{author}{N.~Tuchinda},
    \bibinfo{author}{C.~A. Schuh},
    \newblock \bibinfo{title}{Overview: The spectral model of grain boundary
        segregation},
    \newblock \bibinfo{journal}{Acta Materialia} \bibinfo{volume}{313}
    (\bibinfo{year}{2026}) \bibinfo{pages}{122109}.
    \bibitem[{Pelleg(1966)}]{Pelleg:1966aa}
    \bibinfo{author}{J.~Pelleg},
    \newblock \bibinfo{title}{On the relation between diffusion coefficients and
        grain boundary energy},
    \newblock \bibinfo{journal}{The Philosophical Magazine: A Journal of
        Theoretical Experimental and Applied Physics} \bibinfo{volume}{14}
    (\bibinfo{year}{1966}) \bibinfo{pages}{595--601}.
    \bibitem[{Guiraldenq(1975)}]{GUIRALDENQ}
    \bibinfo{author}{P.~Guiraldenq},
    \newblock \bibinfo{title}{Diffusion intergranulaire et energie des joints de
        grains},
    \newblock \bibinfo{journal}{J. Phys. Colloques} \bibinfo{volume}{36}
    (\bibinfo{year}{1975}) \bibinfo{pages}{C4--201--C4--211}.
    \bibitem[{Gupta(1977)}]{Gupta:1977aa}
    \bibinfo{author}{D.~Gupta},
    \newblock \bibinfo{title}{Influence of solute segregation on grain-boundary
        energy and self-diffusion},
    \newblock \bibinfo{journal}{Metallurgical Transactions A} \bibinfo{volume}{8}
    (\bibinfo{year}{1977}) \bibinfo{pages}{1431--1438}.
    \bibitem[{Li et~al.(2023)Li, Lu, Divinski, and Vitos}]{Li:2023ab}
    \bibinfo{author}{C.~Li}, \bibinfo{author}{S.~Lu},
    \bibinfo{author}{S.~Divinski}, \bibinfo{author}{L.~Vitos},
    \newblock \bibinfo{title}{Theoretical and experimental grain boundary energies
        in body-centered cubic metals},
    \newblock \bibinfo{journal}{Acta Materialia} \bibinfo{volume}{255}
    (\bibinfo{year}{2023}) \bibinfo{pages}{119074}.
    \bibitem[{Zhu et~al.(2018)Zhu, Samanta, Li, Rudd, and Frolov}]{Zhu:2018aa}
    \bibinfo{author}{Q.~Zhu}, \bibinfo{author}{A.~Samanta},
    \bibinfo{author}{B.~Li}, \bibinfo{author}{R.~E. Rudd},
    \bibinfo{author}{T.~Frolov},
    \newblock \bibinfo{title}{Predicting phase behavior of grain boundaries with
        evolutionary search and machine learning},
    \newblock \bibinfo{journal}{Nature Communications} \bibinfo{volume}{9}
    (\bibinfo{year}{2018}) \bibinfo{pages}{467}.
    \bibitem[{Frolov et~al.(2026)Frolov, Neugebauer, and
        Mishin}]{MRS-Bulletin-GB-phases}
    \bibinfo{author}{T.~Frolov}, \bibinfo{author}{J.~Neugebauer},
    \bibinfo{author}{Y.~Mishin},
    \newblock \bibinfo{title}{Thermodynamics of grain-boundary phases},
    \newblock \bibinfo{journal}{MRS Bulletin} \bibinfo{volume}{51}
    (\bibinfo{year}{2026}) \bibinfo{pages}{1--15}.
    \bibitem[{Frolov and Mishin(2012)}]{Frolov2012b}
    \bibinfo{author}{T.~Frolov}, \bibinfo{author}{Y.~Mishin},
    \newblock \bibinfo{title}{Thermodynamics of coherent interfaces under
        mechanical stresses. {II. Application} to atomistic simulation of grain
        boundaries},
    \newblock \bibinfo{journal}{Phys. Rev. B} \bibinfo{volume}{85}
    (\bibinfo{year}{2012}) \bibinfo{pages}{224107}.
    \bibitem[{Chen et~al.(2003)Chen, Tiwari, Iijima, and Yamauchi}]{Chen:2003aa}
    \bibinfo{author}{T.-F. Chen}, \bibinfo{author}{G.~P. Tiwari},
    \bibinfo{author}{Y.~Iijima}, \bibinfo{author}{K.~Yamauchi},
    \newblock \bibinfo{title}{Volume and grain boundary diffusion of chromium in
        {Ni}-base {Ni-Cr-Fe} alloys},
    \newblock \bibinfo{journal}{Materials Transactions} \bibinfo{volume}{44}
    (\bibinfo{year}{2003}) \bibinfo{pages}{40--46}.
    \bibitem[{Divinski et~al.(2010)Divinski, Reglitz, and Wilde}]{Divinski:2010aa}
    \bibinfo{author}{S.~V. Divinski}, \bibinfo{author}{G.~Reglitz},
    \bibinfo{author}{G.~Wilde},
    \newblock \bibinfo{title}{Grain boundary self-diffusion in polycrystalline
        nickel of different purity levels},
    \newblock \bibinfo{journal}{Acta Materialia} \bibinfo{volume}{58}
    (\bibinfo{year}{2010}) \bibinfo{pages}{386--395}.
    \bibitem[{Prokoshkina et~al.(2013)Prokoshkina, Esin, Wilde, and
        Divinski}]{Prokoshkina:2013aa}
    \bibinfo{author}{D.~Prokoshkina}, \bibinfo{author}{V.~A. Esin},
    \bibinfo{author}{G.~Wilde}, \bibinfo{author}{S.~V. Divinski},
    \newblock \bibinfo{title}{Grain boundary width, energy and self-diffusion in
        nickel: {Effect} of material purity},
    \newblock \bibinfo{journal}{Acta Materialia} \bibinfo{volume}{61}
    (\bibinfo{year}{2013}) \bibinfo{pages}{5188--5197}.
    \bibitem[{Tuchinda and Schuh(2024)}]{Tuchinda:2024aa}
    \bibinfo{author}{N.~Tuchinda}, \bibinfo{author}{C.~A. Schuh},
    \newblock \bibinfo{title}{Triple junction excess energy in polycrystalline
        metals},
    \newblock \bibinfo{journal}{Acta Materialia} \bibinfo{volume}{279}
    (\bibinfo{year}{2024}) \bibinfo{pages}{120274}.
    \bibitem[{Lin et~al.(2017)Lin, Laughlin, and Zhu}]{Lin:2017aa}
    \bibinfo{author}{Y.~Lin}, \bibinfo{author}{D.~E. Laughlin},
    \bibinfo{author}{J.~Zhu},
    \newblock \bibinfo{title}{A study of the determination of grain boundary
        diffusivity and energy through the thermally grown oxide ridges on a
        {Fe-22Cr} alloy surface},
    \newblock \bibinfo{journal}{Philosophical Magazine} \bibinfo{volume}{97}
    (\bibinfo{year}{2017}) \bibinfo{pages}{535--548}.
    \bibitem[{Gibbs(1948)}]{Gibbs}
    \bibinfo{author}{J.~W. Gibbs}, \bibinfo{title}{The collected works of J. W.
        Gibbs}, volume~\bibinfo{volume}{1}, \bibinfo{publisher}{Yale University
        Press}, \bibinfo{address}{New Haven}, \bibinfo{year}{1948}.
    \bibitem[{McLean(1957)}]{McLean}
    \bibinfo{author}{D.~McLean}, \bibinfo{title}{Grain Boundaries in Metals},
    \bibinfo{publisher}{Clarendon Press}, \bibinfo{address}{Oxford},
    \bibinfo{year}{1957}.
    \bibitem[{Vaidya et~al.(2017)Vaidya, Pradeep, Murty, Wilde, and
        Divinski}]{Vaidya:2017aa}
    \bibinfo{author}{M.~Vaidya}, \bibinfo{author}{K.~G. Pradeep},
    \bibinfo{author}{B.~S. Murty}, \bibinfo{author}{G.~Wilde},
    \bibinfo{author}{S.~V. Divinski},
    \newblock \bibinfo{title}{Radioactive isotopes reveal a non sluggish kinetics
        of grain boundary diffusion in high entropy alloys.},
    \newblock \bibinfo{journal}{Sci Rep} \bibinfo{volume}{7} (\bibinfo{year}{2017})
    \bibinfo{pages}{12293}.
    \bibitem[{Glienke et~al.(2020)Glienke, Vaidya, Gururaj, Daum, Tas, Rogal,
        Pradeep, Divinski, and Wilde}]{Glienke:2020aa}
    \bibinfo{author}{M.~Glienke}, \bibinfo{author}{M.~Vaidya},
    \bibinfo{author}{K.~Gururaj}, \bibinfo{author}{L.~Daum},
    \bibinfo{author}{B.~Tas}, \bibinfo{author}{L.~Rogal}, \bibinfo{author}{K.~G.
        Pradeep}, \bibinfo{author}{S.~V. Divinski}, \bibinfo{author}{G.~Wilde},
    \newblock \bibinfo{title}{Grain boundary diffusion in {CoCrFeMnNi} high entropy
        alloy: {Kinetic} hints towards a phase decomposition},
    \newblock \bibinfo{journal}{Acta Materialia} \bibinfo{volume}{195}
    (\bibinfo{year}{2020}) \bibinfo{pages}{304--316}.
    \bibitem[{Choi et~al.(2024)Choi, da~Silva~Pinto, Yang, Yu, Lee, Luckabauer,
        Wilde, and Divinski}]{Choi:2024aa}
    \bibinfo{author}{N.~Choi}, \bibinfo{author}{M.~da~Silva~Pinto},
    \bibinfo{author}{S.~Yang}, \bibinfo{author}{J.~H. Yu}, \bibinfo{author}{J.-S.
        Lee}, \bibinfo{author}{M.~Luckabauer}, \bibinfo{author}{G.~Wilde},
    \bibinfo{author}{S.~V. Divinski},
    \newblock \bibinfo{title}{Grain boundary diffusion in additively manufactured
        {CoCrFeMnNi} high-entropy alloys: Impact of non-equilibrium state,
        temperature and relaxation},
    \newblock \bibinfo{journal}{Materialia} \bibinfo{volume}{38}
    (\bibinfo{year}{2024}) \bibinfo{pages}{102228}.
    \bibitem[{Yadav et~al.(2026)Yadav, Mohan~Muralikrishna, Vaidya, Wilde, and
        Divinski}]{Yadav:2026aa}
    \bibinfo{author}{B.~Yadav}, \bibinfo{author}{G.~Mohan~Muralikrishna},
    \bibinfo{author}{M.~Vaidya}, \bibinfo{author}{G.~Wilde},
    \bibinfo{author}{S.~V. Divinski},
    \newblock \bibinfo{title}{Grain boundary diffusion in compositionally complex
        alloys: {A} comprehensive review},
    \newblock \bibinfo{journal}{International Materials Reviews}
    \bibinfo{volume}{71} (\bibinfo{year}{2026}) \bibinfo{pages}{129--177}.
    \bibitem[{Birnbaum et~al.(1988)Birnbaum, Landa, and Kimura}]{Birnmaun-1988}
    \bibinfo{author}{H.~Birnbaum}, \bibinfo{author}{B.~Landa},
    \bibinfo{author}{A.~Kimura},
    \newblock \bibinfo{title}{Hydrogen segregation and diffusion at grain
        boundaries},
    \newblock \bibinfo{journal}{Journal de Physique Colloques} \bibinfo{volume}{49}
    (\bibinfo{year}{1988}) \bibinfo{pages}{C5--349--C5--401}.
    \bibitem[{Yao and Cahoon(1991)}]{Yao:1991aa}
    \bibinfo{author}{J.~Yao}, \bibinfo{author}{J.~R. Cahoon},
    \newblock \bibinfo{title}{Experimental studies of grain boundary diffusion of
        hydrogen in metals},
    \newblock \bibinfo{journal}{Acta Metallurgica et Materialia}
    \bibinfo{volume}{39} (\bibinfo{year}{1991}) \bibinfo{pages}{119--126}.
    \bibitem[{Harris and Latanision(1991)}]{Harris:1991aa}
    \bibinfo{author}{T.~M. Harris}, \bibinfo{author}{M.~Latanision},
    \newblock \bibinfo{title}{Grain boundary diffusion of hydrogen in nickel},
    \newblock \bibinfo{journal}{Metallurgical Transactions A} \bibinfo{volume}{22}
    (\bibinfo{year}{1991}) \bibinfo{pages}{351--355}.
    \bibitem[{Brass and Chanfreau(1996)}]{Brass:1996aa}
    \bibinfo{author}{A.~M. Brass}, \bibinfo{author}{A.~Chanfreau},
    \newblock \bibinfo{title}{Accelerated diffusion of hydrogen along grain
        boundaries in nickel},
    \newblock \bibinfo{journal}{Acta Materialia} \bibinfo{volume}{44}
    (\bibinfo{year}{1996}) \bibinfo{pages}{3823--3831}.
    \bibitem[{Pedersen and J\'{o}nsson(2009)}]{Pedersen2009}
    \bibinfo{author}{A.~Pedersen}, \bibinfo{author}{H.~J\'{o}nsson},
    \newblock \bibinfo{title}{Simulations of hydrogen diffusion at grain boundaries
        in aluminum},
    \newblock \bibinfo{journal}{Acta Mater.} \bibinfo{volume}{57}
    (\bibinfo{year}{2009}) \bibinfo{pages}{4036}.
    \bibitem[{Oudriss et~al.(2012)Oudriss, Creus, Bouhattate, Conforto, Berziou,
        Savall, and Feaugas}]{Oudriss:2012aa}
    \bibinfo{author}{A.~Oudriss}, \bibinfo{author}{J.~Creus},
    \bibinfo{author}{J.~Bouhattate}, \bibinfo{author}{E.~Conforto},
    \bibinfo{author}{C.~Berziou}, \bibinfo{author}{C.~Savall},
    \bibinfo{author}{X.~Feaugas},
    \newblock \bibinfo{title}{Grain size and grain-boundary effects on diffusion
        and trapping of hydrogen in pure nickel},
    \newblock \bibinfo{journal}{Acta Materialia} \bibinfo{volume}{60}
    (\bibinfo{year}{2012}) \bibinfo{pages}{6814--6828}.
    \bibitem[{Zhou et~al.(2019)Zhou, Mousseau, and Song}]{Zhou:2019aa}
    \bibinfo{author}{X.~Zhou}, \bibinfo{author}{N.~Mousseau},
    \bibinfo{author}{J.~Song},
    \newblock \bibinfo{title}{Is hydrogen diffusion along grain boundaries fast or
        slow? {Atomistic} origin and mechanistic modeling},
    \newblock \bibinfo{journal}{Physical Review Letters} \bibinfo{volume}{122}
    (\bibinfo{year}{2019}) \bibinfo{pages}{215501}.
    \bibitem[{Page et~al.(2021)Page, Varela, Johnson, Fullwood, and
        Homer}]{Page:2021aa}
    \bibinfo{author}{D.~E. Page}, \bibinfo{author}{K.~F. Varela},
    \bibinfo{author}{O.~K. Johnson}, \bibinfo{author}{D.~T. Fullwood},
    \bibinfo{author}{E.~R. Homer},
    \newblock \bibinfo{title}{Measuring simulated hydrogen diffusion in symmetric
        tilt nickel grain boundaries and examining the relevance of the {Borisov}
        relationship for individual boundary diffusion},
    \newblock \bibinfo{journal}{Acta Materialia} \bibinfo{volume}{212}
    (\bibinfo{year}{2021}) \bibinfo{pages}{116882}.
    \bibitem[{Wang et~al.(2022)Wang, Wang, Li, Liu, Zhao, and Xu}]{Wang:2022ac}
    \bibinfo{author}{Y.~Wang}, \bibinfo{author}{H.~Wang}, \bibinfo{author}{L.~Li},
    \bibinfo{author}{J.~Liu}, \bibinfo{author}{P.~Zhao}, \bibinfo{author}{Z.~Xu},
    \newblock \bibinfo{title}{The effect of symmetrically tilt grain boundary of
        aluminum on hydrogen diffusion},
    \newblock \bibinfo{journal}{Metals} \bibinfo{volume}{12} (\bibinfo{year}{2022})
    \bibinfo{pages}{345}.
    \bibitem[{Mishin(2015)}]{Mishin:2015ac}
    \bibinfo{author}{Y.~Mishin},
    \newblock \bibinfo{title}{An atomistic view of grain boundary diffusion},
    \newblock \bibinfo{journal}{Defect and Diffusion Forum} \bibinfo{volume}{363}
    (\bibinfo{year}{2015}) \bibinfo{pages}{1--11 (arXiv:1904.10756)}.
    \bibitem[{Suzuki and Mishin(2005)}]{Suzuki05a}
    \bibinfo{author}{A.~Suzuki}, \bibinfo{author}{Y.~Mishin},
    \newblock \bibinfo{title}{Atomic mechanisms of grain boundary diffusion: Low
        versus high temperatures},
    \newblock \bibinfo{journal}{J. Mater. Sci.} \bibinfo{volume}{40}
    (\bibinfo{year}{2005}) \bibinfo{pages}{3155--3161}.
    \bibitem[{Chesser et~al.(2024)Chesser, Koju, and Mishin}]{Chesser:2024aa}
    \bibinfo{author}{I.~Chesser}, \bibinfo{author}{R.~K. Koju},
    \bibinfo{author}{Y.~Mishin},
    \newblock \bibinfo{title}{Atomic-level mechanisms of short-circuit diffusion in
        materials},
    \newblock \bibinfo{journal}{International journal of materials research}
    \bibinfo{volume}{115} (\bibinfo{year}{2024}) \bibinfo{pages}{85--105}.
    \bibitem[{J{\'o}nsson et~al.(1998)J{\'o}nsson, Mills, and Jacobsen}]{Jonsson98}
    \bibinfo{author}{H.~J{\'o}nsson}, \bibinfo{author}{G.~Mills},
    \bibinfo{author}{K.~W. Jacobsen},
    \newblock \bibinfo{title}{Nudged elastic band method for finding minimum energy
        paths of transitions},
    \newblock in: \bibinfo{editor}{B.~J. Berne}, \bibinfo{editor}{G.~Ciccotti},
    \bibinfo{editor}{D.~F. Coker} (Eds.), \bibinfo{booktitle}{Classical and
        Quantum Dynamics in Condensed Phase Simulations}, \bibinfo{publisher}{World
        Scientific, Singapore}, \bibinfo{year}{1998}, p.~\bibinfo{pages}{1}.
    \bibitem[{Henkelman and Jonsson(2000)}]{HenkelmanJ00}
    \bibinfo{author}{G.~Henkelman}, \bibinfo{author}{H.~Jonsson},
    \newblock \bibinfo{title}{Improved tangent estimate in the nudged elastic band
        method for finding minimum energy paths and saddle points},
    \newblock \bibinfo{journal}{J. Chem. Phys.} \bibinfo{volume}{113}
    (\bibinfo{year}{2000}) \bibinfo{pages}{9978--9985}.
    \bibitem[{Urazaliev et~al.(2024)Urazaliev, Stupak, and
        Popov}]{Urazaliev:2024aa}
    \bibinfo{author}{M.~Urazaliev}, \bibinfo{author}{M.~Stupak},
    \bibinfo{author}{V.~Popov},
    \newblock \bibinfo{title}{Calculation of {GB} energies and grain-boundary
        self-diffusion in nickel and verification of {Borisov} relations for various
        symmetric tilt grain boundaries},
    \newblock \bibinfo{journal}{Journal of Phase Equilibria and Diffusion}
    \bibinfo{volume}{45} (\bibinfo{year}{2024}) \bibinfo{pages}{384--396}.
    \bibitem[{Borisov et~al.(1962)Borisov, Golikov, and
        Shcherbedinskiy}]{Borisov-1962}
    \bibinfo{author}{V.~T. Borisov}, \bibinfo{author}{V.~M. Golikov},
    \bibinfo{author}{G.~V. Shcherbedinskiy}, \bibinfo{title}{Problems of
        Metallography and Metal Physics}, volume~\bibinfo{volume}{7},
    \bibinfo{publisher}{Metallurgizdat}, \bibinfo{year}{1962}, p.
    \bibinfo{pages}{501}.
    
\end{thebibliography}

\newpage\clearpage{}

\section*{Appendix}

In this appendix, we examine the equation 
\begin{equation}
\dfrac{D^{\prime}}{D}=\left(\dfrac{\tau}{\tau^{\prime}}\right)^{\alpha}\label{eq:100}
\end{equation}
derived by Borisov et al.~\citep{Borisov,Borisov-1962} for self-diffusion.
Here, $D$ and $D^{\prime}$ are the diffusion coefficients in the
lattice and in the GB, and $\tau$ and $\tau^{\prime}$ are the respective
residence times of the atoms. We first consider self-diffusion by
the vacancy mechanism.

Generally, the self-diffusion coefficient by the vacancy mechanism
is given by Eq.(\ref{eq:16}), from which we obtain 
\begin{equation}
\dfrac{D^{\prime}}{D}=\dfrac{f^{\prime}\xi^{\prime}l^{\prime2}}{f\xi l^{2}}\dfrac{c_{v}^{\prime}}{c_{v}}\dfrac{w^{\prime}}{w},\label{eq:101}
\end{equation}
where $w$ and $w^{\prime}$ are the respective vacancy-atom exchange
rates. Borisov et al.~\citep{Borisov,Borisov-1962} adopted an approximate
form of this equation, which in our notation reads: 
\begin{equation}
\dfrac{D^{\prime}}{D}=\dfrac{c_{v}^{\prime}}{c_{v}}\dfrac{\tau}{\tau^{\prime}}.\label{eq:102}
\end{equation}
They neglected the structural pre-factors, assuming that the environments
of atoms in the lattice and in the GB are identical. They also assumed
that the residence times were proportional to the vacancy-atom exchange
rates $w$ and $w^{\prime}$. This assumption is valid if the vacancies
have the same number of neighbors in the GB as in the lattice and
if the exchange rate is the same for all neighbors.

To calculate the ratio $c_{v}^{\prime}/c_{v}$, Borisov et al.~\citep{Borisov,Borisov-1962}
invoked the material balance condition, which they wrote in the form
\begin{equation}
\dfrac{c_{v}^{\prime}}{c_{v}}=\dfrac{\tau}{\tau^{\prime}}.\label{eq:103}
\end{equation}
A material balance equation equates two opposite fluxes: (1) the atomic
flux from the lattice into the GB, which they consider proportional
to $c_{v}^{\prime}w$ (a lattice atom can only jump into the GB if
it can fill a vacant GB site), and (2) the atomic flux from the GB
to the lattice, which they consider proportional to $c_{v}w^{\prime}$
(a GB atom can only jump into the lattice if it can fill a vacant
lattice site). Thus, 
\begin{equation}
c_{v}^{\prime}w=c_{v}w^{\prime}.\label{eq:104}
\end{equation}
Eq.(\ref{eq:103}) then follows by replacing $w^{\prime}/w$ with
$\tau/\tau^{\prime}$.

The problem with the above derivation is that the ratio $\tau/\tau^{\prime}$
appearing in Eq.(\ref{eq:102}) is \emph{not} the same as the ratio
$\tau/\tau^{\prime}$ appearing in Eq.(\ref{eq:103}). In other words,
the frequency ratios $w^{\prime}/w$ in the two equations are different
in their physical meaning and magnitude. The ratio $w^{\prime}/w$
appearing in Eq.(\ref{eq:102}) is one between two \emph{independent}
transitions: a vacancy-atom exchange in the lattice and a vacancy-atom
exchange in the GB. According to equations (\ref{eq:4}) and (\ref{eq:10})
in the main text, 
\begin{equation}
w=\dfrac{kT}{h}\exp\left(-\dfrac{G_{N-1}^{(v)*}-G_{N-1}^{(v)}}{kT}\right),\label{eq:105}
\end{equation}
\begin{equation}
w^{\prime}=\dfrac{kT}{h}\exp\left(-\dfrac{G_{N^{\prime}-1}^{(v)\prime*}-G_{N^{\prime}-1}^{(v)\prime}}{kT}\right).\label{eq:106}
\end{equation}
The notation is explained in the main text. It is convenient to rewrite
these equations in terms of the free energies of the respective activated
complex, $\hat{g}$ and $\hat{g}^{\prime}$, relative to a common
reference. These are given by equations (\ref{eq:36}) and (\ref{eq:37})
with $n=1$. Using those equations, we obtain 
\begin{equation}
w=\dfrac{kT}{h}\exp\left(-\dfrac{\hat{g}-\left[G_{N-1}^{(v)}-\mu_{0}N\right]-2\mu_{0}}{kT}\right),\label{eq:107}
\end{equation}
\begin{equation}
w^{\prime}=\dfrac{kT}{h}\exp\left(-\dfrac{\hat{g}^{\prime}-\left[G_{N^{\prime}-1}^{(v)\prime}-\mu_{0}^{\prime}N^{\prime}\right]-2\mu_{0}^{\prime}}{kT}\right).\label{eq:108}
\end{equation}
The terms in the square brackets are the non-configurational parts
of the diffusion potentials of the vacancies. For the ratio of the
two rates, we have 
\begin{equation}
\dfrac{w^{\prime}}{w}=\exp\left(-\dfrac{\hat{g}^{\prime}-\hat{g}}{kT}\right)\exp\left(\dfrac{\left[G_{N^{\prime}-1}^{(v)\prime}-\mu_{0}^{\prime}N^{\prime}\right]-\left[G_{N-1}^{(v)}-\mu_{0}N\right]}{kT}\right)\exp\left(\dfrac{2(\mu_{0}^{\prime}-\mu_{0})}{kT}\right).\label{eq:109}
\end{equation}
Note that this equation combines thermodynamics and kinetics. The
first exponential factor depends on the free energies of the activated
complexes, $\hat{g}$ and $\hat{g}^{\prime}$, which are generally
not equal. Figure \ref{fig:1} is a schematic illustration of the
two activated complexes. The remaining exponential factors are thermodynamic
in nature and depend, respectively, on the diffusion potentials of
the vacancies and the chemical potentials of the atoms in the GB and
in the lattice.

The situation is different for atomic jumps \emph{between} the GB
and the lattice. Let us denote the respective jump frequencies $w_{t}$
and $w_{t}^{\prime}$ (Fig. \ref{fig:1}). The mass balance equation
should be written in the form 
\begin{equation}
c_{v}^{\prime}w_{t}=c_{v}w_{t}^{\prime},\label{eq:110}
\end{equation}
from which 
\begin{equation}
\dfrac{c_{v}^{\prime}}{c_{v}}=\dfrac{w_{t}^{\prime}}{w_{t}}.\label{eq:115}
\end{equation}
Clearly, $w_{t}^{\prime}/w_{t}$ must be different from $w^{\prime}/w$
because the ratio $c_{v}^{\prime}/c_{v}$ is a thermodynamic property
that cannot depend on any transition barriers, such as those appearing
in Eq.(\ref{eq:109}).

To find $w_{t}$ and $w_{t}^{\prime}$, we need to consider a compound
system composed of the GB and a lattice region. This will allow us
to consider atomic jumps between them. Let the numbers of atoms in
the two regions in the vacancy-free state be, respectively, $N^{\prime}$
and $N$.

Both forward and backward jumps between the GB and lattice sites occur
through the \emph{same} activated complex containing a vacancy at
the interface between the two regions. Let the total free energy of
the system in the activated state be $G_{N+N^{\prime}-1}^{(v)*}$,
and let the total free energy before the jump be $G_{N+N^{\prime}-1}^{(v)}$
if the vacancy originates from the lattice and $G_{N+N^{\prime}-1}^{(v)\prime}$
otherwise. Then the atomic jump rates are, respectively, 
\begin{equation}
w_{t}=\dfrac{kT}{h}\exp\left(-\dfrac{G_{N+N^{\prime}-1}^{(v)*}-G_{N+N^{\prime}-1}^{(v)\prime}}{kT}\right),\label{eq:117}
\end{equation}
\begin{equation}
w_{t}^{\prime}=\dfrac{kT}{h}\exp\left(-\dfrac{G_{N+N^{\prime}-1}^{(v)*}-G_{N+N^{\prime}-1}^{(v)}}{kT}\right),\label{eq:116}
\end{equation}
from which 
\begin{equation}
\dfrac{w_{t}^{\prime}}{w_{t}}=\exp\left(-\dfrac{G_{N+N^{\prime}-1}^{(v)\prime}-G_{N+N^{\prime}-1}^{(v)}}{kT}\right).\label{eq:118}
\end{equation}
As expected, $w^{\prime}/w$ depends only on the free energies of
the vacancy in the two regions and not on the activated state (note
that $G_{N+N^{\prime}-1}^{(v)*}$ cancels out).

To be more specific, we can represent $G_{N+N^{\prime}-1}^{(v)}$
by 
\begin{equation}
G_{N+N^{\prime}-1}^{(v)}=N^{\prime}\mu_{0}^{\prime}+G_{N-1}^{(v)},\label{eq:119}
\end{equation}
where the first term is the free energy of the vacancy-free GB and
the second term is the free energy of the lattice region with a vacancy.
Similarly, 
\begin{equation}
G_{N+N^{\prime}-1}^{(v)\prime}=N\mu_{0}+G_{N^{\prime}-1}^{\prime(v)},\label{eq:120}
\end{equation}
where the first term is the free energy of the vacancy-free lattice
and the second term is the free energy of the GB with a vacancy. Inserting
equations (\ref{eq:119}) and (\ref{eq:120}) into Eq.(\ref{eq:118}),
we have 
\begin{equation}
\dfrac{w_{t}^{\prime}}{w_{t}}=\exp\left(-\dfrac{\left[G_{N^{\prime}-1}^{(v)\prime}-N^{\prime}\mu_{0}^{\prime}\right]-\left[G_{N-1}^{(v)}-N\mu_{0}\right]}{kT}\right).\label{eq:121}
\end{equation}
This expression exactly matches Eq.(\ref{eq:38}) for the vacancy
concentration ratio $c_{v}^{\prime}/c_{v}$ derived in section \ref{subsec:Self-diffusion-vacancy}
of the main text, confirming the correctness of Eq.(\ref{eq:115}).

Combining equations (\ref{eq:109}) and (\ref{eq:121}), we obtain
\begin{equation}
\dfrac{D^{\prime}}{D}=\dfrac{w^{\prime}}{w}\dfrac{w_{t}^{\prime}}{w_{t}}=\exp\left(-\dfrac{\hat{g}^{\prime}-\hat{g}}{kT}\right)\exp\left(\dfrac{2(\mu_{0}^{\prime}-\mu_{0})}{kT}\right),\label{eq:122}
\end{equation}
which recovers Eq.(\ref{eq:44}) with $n=1$ if we neglect the structural
factors and jump correlation effects.

It is interesting to express the atomic jump rates, $w_{t}$ and $w_{t}^{\prime}$,
in terms of the free energy $\hat{g}_{t}$ of the activated complex
(Fig. \ref{fig:1}). Assuming that the latter includes one atom from
each region ($n=2$), we have 
\begin{equation}
\hat{g}_{t}=G_{N+N^{\prime}-1}^{(v)*}-(N^{\prime}-1)\mu_{0}^{\prime}-(N-1)\mu_{0}.\label{eq:123}
\end{equation}
Inserting this expression into equations (\ref{eq:117}) and (\ref{eq:116}),
we obtain 
\begin{equation}
w_{t}^{\prime}=\dfrac{kT}{h}\exp\left(-\dfrac{\hat{g}_{t}-\left[G_{N-1}^{(v)}-N\mu_{0}\right]-(\mu_{0}^{\prime}+\mu_{0})}{kT}\right),\label{eq:124}
\end{equation}
\begin{equation}
w_{t}=\dfrac{kT}{h}\exp\left(-\dfrac{\hat{g}_{t}-\left[G_{N^{\prime}-1}^{\prime(v)}-N^{\prime}\mu_{0}^{\prime}\right]-(\mu_{0}^{\prime}+\mu_{0})}{kT}\right).\label{eq:125}
\end{equation}
Compare these equations with equations (\ref{eq:107}) and (\ref{eq:108})
for $w$ and $w^{\prime}$. The different free energies $\hat{g}$
and $\hat{g}^{\prime}$ are replaced by the single free energy $\hat{g}_{t}$,
and the terms $2\mu_{0}$ and $2\mu_{0}^{\prime}$ are replaced by
the sum $\mu_{0}+\mu_{0}^{\prime}$. These changes reflect the difference
between the two types of atomic jumps, with $w$ and $w^{\prime}$
representing the rates of different and independent jumps and $w_{t}$
and $w_{t}^{\prime}$ representing the rates of the \emph{same} jumps
in the forward and backward directions.

In summary, Eq.(\ref{eq:100}) derived by Borisov et al.~\citep{Borisov,Borisov-1962}
for self-diffusion by the vacancy mechanism is conceptually invalid.
It confuses different types of atomic jumps. However, under certain
approximations and assuming that $\hat{g}^{\prime}=\hat{g}$ holds
exactly, Eq.(\ref{eq:100}) leads to the same expression for $D^{\prime}/D$
as our calculation under the same approximations.

Next, consider diffusion of impurity atoms by the direct interstitial
mechanism. As shown in the main text, in this case the diffusivity
ratio $D^{\prime}/D$ does not depend on the GB free energy and is
determined solely by the GB segregation factor. If the GB is enriched
in the impurity, then GB diffusion of the impurity atoms is slower
than lattice diffusion. The effect is illustrated in Figure \ref{fig:2}.
The GB segregation factor $g_{s}$ is given by Eq.(\ref{eq:52-1}),
which can be written in the form 
\begin{equation}
g_{s}=\tilde{\mu}_{I}^{\prime}-\tilde{\mu}_{I},\label{eq:126}
\end{equation}
where 
\begin{equation}
\tilde{\mu}_{I}=G_{N}^{(I)}-N\mu_{0}\label{eq:127}
\end{equation}
and 
\begin{equation}
\tilde{\mu}_{I}^{\prime}=G_{N\prime}^{(I)\prime}-N^{\prime}\mu_{0}^{\prime}\label{eq:128}
\end{equation}
are the non-configurational parts of the chemical potentials of the
impurity in the lattice and in the GB, respectively. Equations (\ref{eq:55})
and (\ref{eq:56}) show that the respective free energies of the activated
complexes are 
\begin{equation}
\hat{g}=g^{*}+\tilde{\mu}_{I}\label{eq:129}
\end{equation}
and 
\begin{equation}
\hat{g}^{\prime}=g^{*\prime}+\tilde{\mu}_{I}^{\prime}.\label{eq:129-1}
\end{equation}
Here, $g^{*}$ and $g^{*\prime}$ are the respective activation free
energies of diffusion given by 
\begin{equation}
g^{*}=G_{N}^{(I)*}-G_{N}^{(I)}\label{eq:140}
\end{equation}
and 
\begin{equation}
g^{*\prime}=G_{N^{\prime}}^{(I)\prime*}-G_{N^{\prime}}^{(I)\prime}.\label{eq:141}
\end{equation}
If the impurity atoms segregate to the GB, then $g_{s}<0$ and thus
$\tilde{\mu}_{I}^{\prime}<\tilde{\mu}_{I}$. If $\hat{g}\approx\hat{g}^{\prime}$,
as assumed in the model, then it follows from equations (\ref{eq:129})
and (\ref{eq:129-1}) that $g^{*\prime}>g^{*}$. The activation free
energy of GB diffusion is higher than the activation energy of lattice
diffusion. In other words, GB diffusion is slower than lattice diffusion. 
\end{document}